\documentclass[12pt]{article}
\usepackage{amsmath}
\usepackage{graphicx}
\usepackage{enumerate}
\usepackage{natbib}
\usepackage{url} % not crucial - just used below for the URL 
\usepackage[colorlinks=true,allcolors=blue]{hyperref}
\usepackage{url}
\usepackage{graphicx}
\usepackage{amsmath, amssymb}
\usepackage{bbold}
\usepackage{mathbbol}
\usepackage{algorithm}
\usepackage{algorithmic}
\usepackage{enumitem}
\usepackage{float}
\usepackage{xcolor}
\usepackage{bm} % Math packages
\usepackage{setspace}
\usepackage{multirow}
\usepackage{lscape}
\usepackage{rotating}
\usepackage{adjustbox}
\usepackage{booktabs}
\usepackage{ragged2e}

\usepackage{caption}
\usepackage{subcaption}
\usepackage{float}
\def\*#1{\mathbf{#1}}
\def\+#1{\amsmathbb{#1}}
\def\^#1{\mathbb{#1}}
\DeclareSymbolFontAlphabet{\amsmathbb}{AMSb}%

\usepackage{amsthm}

\usepackage[english]{babel}
\usepackage{amsthm}
\usepackage{amssymb}
\newcommand{\blind}{1}

\begin{document}

\def\spacingset#1{\renewcommand{\baselinestretch}%
{#1}\small\normalsize} \spacingset{1}

\newtheorem{corollary}{Corollary}[theorem]
\newtheorem{lemma}[theorem]{Lemma}
%%%%%%%%%%%%%%%%%%%%%%%%%%%%%%%%%%%%%%%%%%%%%%%%%%%%%%%%%%%%%%%%%%%%%%%%%%%%%%

\if1\blind
{
  \title{\bf Functional Time Transformation Model with Applications to Digital Health}
  \author{Rahul Ghosal$^1$,  Marcos Matabuena$^{2}$, Sujit K. Ghosh$^{3}$\\
     $^{1}$ Department of Epidemiology and Biostatistics, University of South Carolina \\
     $^{2}$Department of Biostatistics, Harvard University, Boston, MA 02115, USA\\
$^{3}$ Department of Statistics, North Carolina State University\\
}
  \maketitle
} \fi

\if0\blind
{
  \bigskip
  \bigskip
  \bigskip
  \begin{center}
    {\LARGE\bf Functional time-transformation model}
\end{center}
  \medskip
} \fi

\bigskip
\begin{abstract}
%In the realm of biomedical sciences, often the imperative is to enhance clinical decision-making, by quantifying the risk between a set of covariates and survival or time-to-event outcomes. 
The advent of wearable and sensor technologies now leads to functional predictors which are intrinsically infinite dimensional. While the existing approaches for functional data and survival outcomes lean on the well-established Cox model, the proportional hazard (PH) assumption might not always be suitable in real-world applications. Motivated by physiological signals encountered in digital medicine, we develop a more general and flexible functional time-transformation model for estimating the conditional survival function with both functional and scalar covariates. A partially functional regression model is used to directly model the survival time on the covariates through an unknown monotone transformation and a known error distribution. We use Bernstein polynomials to model the monotone transformation function and the smooth functional coefficients. A sieve method of maximum likelihood is employed for estimation. Numerical simulations illustrate a satisfactory performance of the proposed method in estimation and inference. We demonstrate the application of the proposed model through two case studies involving wearable data  i) Understanding the association between diurnal physical activity pattern and all-cause mortality based on accelerometer data from the National Health and Nutrition Examination Survey (NHANES) 2011-2014 and ii) Modelling Time-to-Hypoglycemia events in a cohort of diabetic patients based on distributional representation of continuous glucose monitoring (CGM) data. The results provide important epidemiological insights into the direct association between survival times and the physiological signals and also exhibit superior predictive performance compared to traditional summary based biomarkers in the CGM study.

\end{abstract}

\noindent%
\vfill
\spacingset{1.8} % DON'T change the spacing!
\section{Introduction}
\label{intro}

Advancements in wearable and sensor technology has resulted into collection of real-time, detailed physiological and behavioural signals tailored to individual users.
%encompassing metrics like heart rate, physical activity (such as steps, activity counts), continuously monitored blood glucose, and many other such biomedical data. 
These continuously collected observations can be treated as functional data \citep{Ramsay05functionaldata}, which are intrinsically infinite dimensional, and can provide for a deeper understanding of the link between human behaviors and health and disease. Functional data analysis (FDA) refers to the branch of statistics analyzing data in the forms of curves or surfaces over a continuous index such as time or space. Functional data analysis has been applied across several areas in biosciences such as  digital health \citep{cui2022fast,ghosal2023variable,ghosal2023functional,matabuena2023distributional,matabuena2023estimating}, genome-wide association studies (GWAS)\citep{wu2010functional,huang2017fgwas}, medical imaging \citep{zipunnikov2011functional,zipunnikov2014longitudinal,li2021functional,koner2024profit}, ecology \citep{ikeda2008application,ghosal2020variable}, intervention studies \citep{ghosal2023shape,coffman2023causal} and many others.

In many biomedical and clinical studies, the objective is to enhance clinical decision-making and interpretation, by quantifying the association between a set of risk-factors and survival or time-to-event outcomes which are subject to censoring. Current biomedical and digital health studies often collect high dimensional physiological signals through wearable or sensor technologies such as physical activity (PA) data via accelerometer, blood glucose concentration using continuous glucose monitoring device (CGM), brain activity using EEG, among many others, which can be treated as Hilbert-space valued functional predictors. The existing approaches for functional data and survival outcomes primarily lean on the well-established Cox model \citep{gellar2015cox,qu2016optimal,kong2018flcrm} and its extensions \citep{cui2021additive,ghosal2023functional}. However the proportional hazard (PH) assumption which forces the hazard ratio to be constant over time might not be suitable in many real-world applications and modelling the hazard itself can face lack of interpretability to the practitioners. Another recently proposed approach in the functional domain include the functional accelerated failure time (AFT) model \citep{liu2024efficient}.

\subsection{Motivating Applications}
\begin{figure}[H]
	\centering
 \begin{subfigure}[b]{0.47\textwidth}
         \centering
        \includegraphics[width=0.8\textwidth,height=0.75\textwidth]{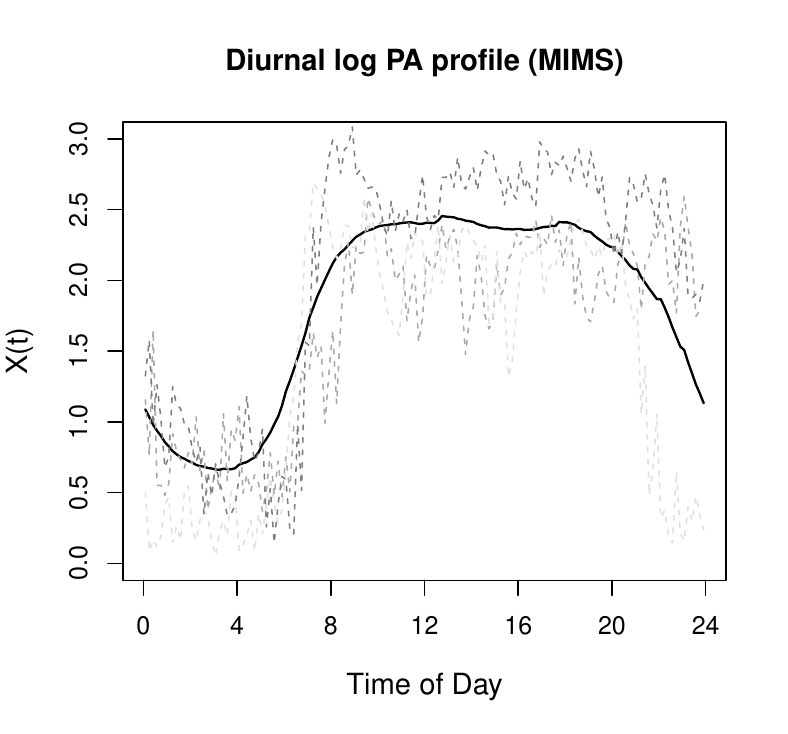}
\caption{Average diurnal PA (MIMS) profile (log-transformed) for NHANES 2011-2014 participants (solid), along with profiles for three randomly chosen participants (dashed).} 
\label{fig:grafactividad1}	
     \end{subfigure}
     %\hfill
     \begin{subfigure}[b]{0.51\textwidth}
\centering\includegraphics[height=0.75\textwidth,width=0.8\textwidth]{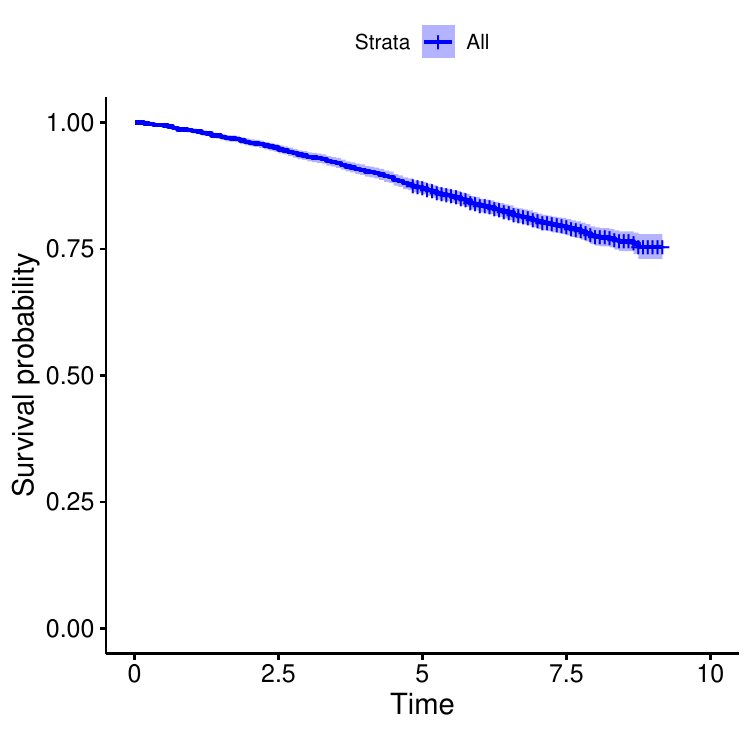}
\caption{Kaplan-Meier marginal survival curve of all-cause mortality in the NHANES 2011-2014.}
         \label{fig:kmage1}
     \end{subfigure}
  \begin{subfigure}[b]{0.47\textwidth}
         \centering
        \includegraphics[width=0.8\textwidth,height=0.75\textwidth]{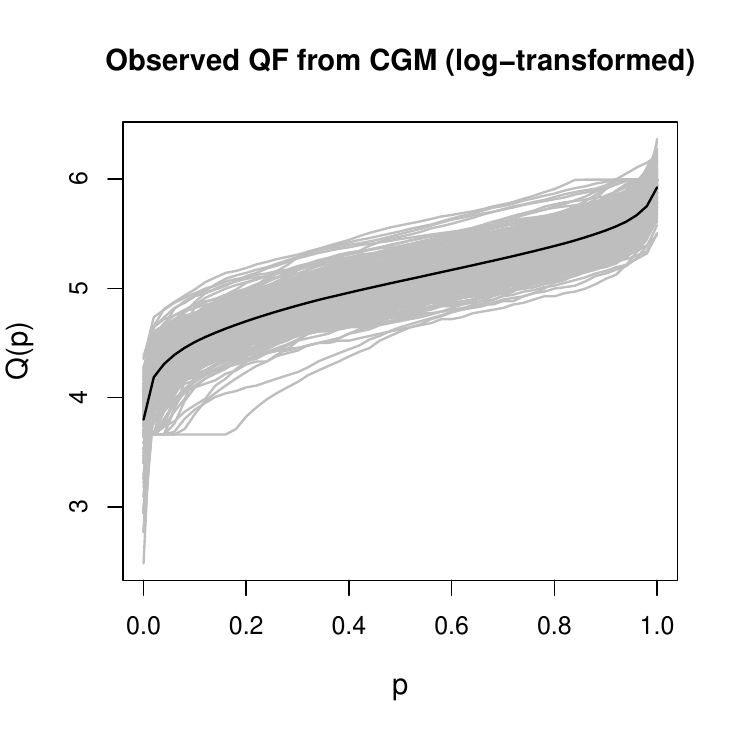}
\caption{Glucodensity profiles of all participants based on subject-specific quantile function representation along with the Wasserstein Barycenter of the profiles in the CGM study (solid).} 
\label{fig:grafactividad2}	
     \end{subfigure}
     %\hfill
     \begin{subfigure}[b]{0.51\textwidth}
\centering\includegraphics[height=0.75\textwidth,width=0.8\textwidth]{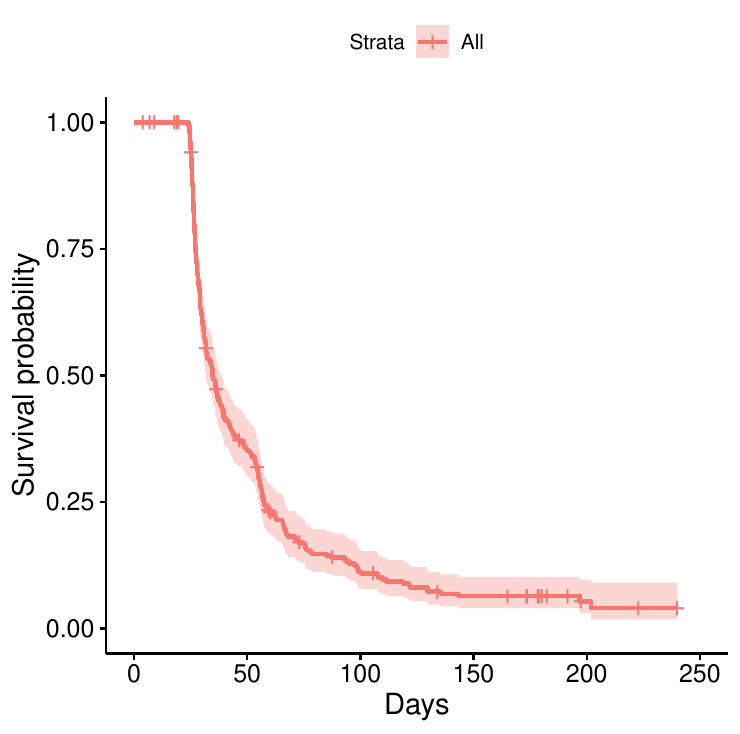}
\caption{Kaplan-Meier plot for time to 65 hypoglycemia events in the CGM study.}
         \label{fig:kmage2}
     \end{subfigure}
     \caption{Functional profiles and Kaplan Meier curves in the wearable applications.}
\end{figure}

In this paper, as motivating applications, we consider two real case studies: (i) Modelling the survival time of older adults (aged 50 years or older)  based on diurnal patterns of objectively measured physical activity data collected using accelerometer in the National Health and Nutrition Examination Survey (NHANES) 2011-2014 and other biological factors such as age, gender and body mass index (BMI). NHANES 2011-2014 reports individuals’ acceleration in Monitor Independent Movement Summary (MIMS) unit \citep{john2019open}. Figure \ref{fig:grafactividad1} displays the average diurnal PA profile across all participants for log-transformed MIMS, along with the PA profile for three sample participants. Figure \ref{fig:kmage1} displays the Kaplan-Meir plot for all-cause mortality through 2019 for the study participants. Previous research in NHANES exploring the relation between PA and all-cause mortality have primarily explored various summary measures of PA \citep{liu2016leisure,patel2019american,smirnova2020predictive}, or have used variants of functional Cox models for modelling the hazard of mortality \citep{cui2021additive} which have mostly been concentrated on the NHANES 2003-06 uni-axial accelerometer data. By directly modelling the survival time on the diurnal PA patterns, we aim to uncover the dynamic association between all-cause mortality and circadian rhythm of PA, which could be useful in designing time-of-day-specific PA interventions.
(ii) We consider modelling time to 65 hypoglycemia events as an adverse outcome in a cohort of Type 1 diabetes mellitus (T1D) patients based on a compositional functional \citep{petersen2016functional} or distributional representation of continuous glucose monitoring (CGM) data collected at baseline. 

Figure \ref{fig:grafactividad2} displays the glucodensity profiles \citep{matabuena2021glucodensities}  of all participants and the Wasserstein Barycenter \citep{bigot2018upper} representing the average quantile function, based on the subject-specific quantile function representations \citep{ghosal2021distributional,matabuena2023distributional} of glucose-concentrations (log-transformed). The glucodensity based marker has been shown to demonstrate improved clinical sensitivity compared to the traditional markers like the time in range metric, HbA1c \citep{matabuena2021glucodensities} in diabetes and they provide more nuanced insight for assessing glucose metabolism. Figure \ref{fig:kmage2} displays the Kaplan-Meir plot for the adverse event ``time-to 65 hypoglycemia" for these study participants.

%\textcolor{red}{Review time transformation models, extensions and motivating application:start with this}.
\subsection{Background and Contributions}
Time-transformation models
\citep{cheng1995analysis,chen2002semiparametric,zeng2006efficient} are a flexible and general class of survival models, which directly model the survival time based on a linear combination of covariates through an unknown monotone transformation $H(\cdot)$ and a known error distribution. The family of transformation models encompasses the popularly used Cox proportional hazards (PH) model, proportional odds (PO) model and accelerated failure time (AFT) models as special cases. Since, the survival time (or it's monotone transformation) is modelled directly in the transformation  models, the covariate effects are directly interpretable in terms of effect on the survival time or survival probability. There exists several estimation approaches in classical transformation models based on generalized estimating equations (GEE) \citep{chen2002semiparametric}, nonparametric maximum likelihood estimation (NPMLE) \citep{zeng2006efficient,zeng2007semiparametric}, and sieve maximum likelihood estimation \citep{mclain2013efficient}.

In this article, we develop a flexible partially functional time-transformation model for estimating the conditional survival function in the presence of both functional and scalar covariates. The proposed model is more general and include both the functional Cox and AFT model as special cases. We use Bernstein polynomials to model the monotone transformation and the smooth functional coefficients, and the sieve method of maximum likelihood \citep{mclain2013efficient,ghosal2023shape} is employed for shape constrained and smooth estimation of the unknown functions. The resulting estimate of conditional survival function is smooth as opposed to the ones generally obtained in the functional cox model \citep{gellar2015cox}, as is the estimate of the monotone transformation function $H(\cdot)$, which facilitates interpretation of the effect of the covariates on survival time.

The key contributions of this paper are i) a partially functional time-transformation model relaxing the PH assumption and moving beyond functional Cox model ii) a computationally efficient sieve maximum likelihood estimation and incorporation of smoothness in the estimated monotone transformation and functional coefficients with Bernstein polynomials and iii) quantifying the association between functional and distributional representations of wearable data and survival time for two case studies;  NHANES 2011-2014 data and the CGM study. In addition, we establish consistency of the proposed estimator under standard regularity conditions. Our empirical analyses using simulations illustrate a satisfactory finite-sample performance the proposed method. The two real case studies provide novel epidemiological insights into the association between the functional biomarkers and the corresponding survival time, which could be important for designing time-of-day or intensity-specific interventions.
%along the different applications examined. 
%talk more about cancer mortality and PA, cite

The rest of this article is organized in the following way. The notations and the modeling framework are introduced in Section \ref{method}. The sieve maximum likelihood estimation method along with other methodological details are presented in Section \ref{est}. Section \ref{simul} reports the empirical analysis using simulation studies. Section \ref{realdat} presents the Real data applications of the proposed method. We conclude with a discussion of the proposed method and some possible extensions of this work in Section \ref{disc}.

\section{Methodology}
\label{method}
Let $T_i$ be the survival time for subject $i$, and $C_i$ the corresponding censoring time for 
$i=1,\ldots,n$. Let us denote $Y_i= \min(T_i,C_i)$ to be the observed survival time for subject $i$ in the presence of right censoring, and $\Delta_i= I(T_i\leq C_i)$ is the event or censoring indicator for subject $i$. Suppose that we observe a random sample $(Y_i,\Delta_i, \*X_i, X_{fi}(\cdot))$, $i=1,\dots, n$, having the same distribution as $(Y,\Delta, \*X, X_{f}(\cdot))$. Conditionally on $(\*X, X_f(\cdot))$, we assume that the survival time $T$ is independent of the censoring time $C$. We observe the functional covariates $X_{fi}(\cdot)$ at the baseline and they are assumed to lie in a real separable Hilbert space, taken to be $L^2[0,1]$ in this paper. The scalar baseline covariates are denoted as $\*X_i=(X_{i1},\ldots,X_{ip})$. We further assume that the functional curves are observed on a dense and regular grid $S= \{s_{1},\ldots,s_{m} \} \subset \mathcal{S}=[0,1] $, although this can be relaxed to accommodate more general scenarios, e.g., functions observed on irregular and sparse domain. 

\subsection{Functional Time Transformation Model}
We posit the functional time transformation model (FTTM), where a monotone transformation of survival time $T$ is additively modelled as linear function of the scalar covariates and a linear functional term of the functional covariate,
\begin{eqnarray}
    H(T)=-\{\bm\beta^T\*X +\int_{0}^{1} X_{f}(s)\beta(s)ds\}+\epsilon 
    = -\{\sum_{j=1}^{p}\beta_jX_j +\int_{0}^{1} X_{f}(s)\beta(s)ds\}+\epsilon \label{eq:ttf1}.
\end{eqnarray}
Here $\epsilon$ is a random variable with a known distribution function $F_{\epsilon}(\cdot)$ and $H(\cdot)$ is an unknown monotone function. Throughout this paper, we assume $H(\cdot)$ to be monotone increasing, without loss of generality. The scalar coefficients $\beta_j$ capture the expected decrease in the transformed survival time $H(T)$ for one unit increase in the covariate $Z_j$. Hence a positive $\beta_j$ indicates a decrease in the expected survival time (since $H(\cdot)$ to be monotone increasing) with increase in $X_j$. Similarly, the functional coefficient $\beta(s)$ capture the expected decrease in the transformed survival time $H(T)$ for one unit increase in the functional covariate $X_f(s)$ at index $s$, keeping all the other covariates fixed. 

The transformation model (\ref{eq:ttf1}) reduces to  proportional hazards model and a proportional odds model when $\epsilon$ follows the extreme-value and logistic distribution respectively. For $H(t)=log(t)$, the transformation model reduces to the well known AFT model. When the residual $\epsilon$ follows a Normal distribution, the model serves as an extension of the usual Box-Cox model. Hence, the transformation model (\ref{eq:ttf1}) provides a generalization to several well known models used in survival and regression analysis. On the other hand, note that, in model (\ref{eq:ttf1}) we are directly modelling the survival time of interest, rather than modelling the hazard or the odds of the event, and hence is more interpretable.

Based on model (\ref{eq:ttf1}), we have $E(H(T)|\*X, X_{f}(\cdot))=-\{\bm\beta^T\*X +\int_{0}^{1} X_{f}(s)\beta(s)ds+\mu_{\epsilon}\}$, hence an approximation of expected survival time is given by $E(T|\*X, X_{f}(\cdot))\approx H^{-1}(-\{\bm\beta^T\*X +\int_{0}^{1} X_{f}(s)\beta(s)ds+\mu_{\epsilon}\})$. The sieve likelihood based estimation procedure, that will be used in this paper ensures a smooth and invertible $H(\cdot)$ function \citep{mclain2013efficient}, where $H^{-1}(\cdot)$ is continuously differentiable and this aids in the interpretation of the coefficients, by associating them to the change in expected survival time $E(T|\*X, X_{f}(\cdot))$. Let us denote the survival function of the  error $\epsilon$ as $\bar{F}_{\epsilon}$ and the density as $f_\epsilon$. The conditional survival function and conditional density of $T$ based on model (\ref{eq:ttf1}) is given by, 
\begin{eqnarray}
  S_T(t|\*X,X_{f}(\cdot))=\bar{F}_{\epsilon}\{ H(t)+\bm\beta^T\*X +\int_{0}^{1} X_{f}(s)\beta(s)ds\}, \label{ttf2}\\
 \bar{F}_{\epsilon}^{-1}\{S_T(t|\*X,X_{f}(\cdot))\}=H(t)+\bm\beta^T\*X +\int_{0}^{1} X_{f}(s)\beta(s)ds, \\
   f_T(t|\*X,X_{f}(\cdot))=H^{'}(t){f}_{\epsilon}\{ H(t)+\bm\beta^T\*X +\int_{0}^{1} X_{f}(s)\beta(s)ds\}.
\end{eqnarray}
The parameters $H(\cdot),\bm\beta,\beta(s)$ are identifiable if $X$ is full rank and $Ke(K^{X_f})=\{0\}$ 
\citep{mclain2013efficient,scheipl2016identifiability}, where $K^{X_f}$ denotes the covariance operator of the functional process $X_f(\cdot)$ and $Ke(C)$ denotes the kernel or the null space of the operator $C$. 
%the eigenfunctions of functional process $X_f(\cdot)$ span $L^2[0,1]$.
For the error $\epsilon$, we specifically consider the class of logarithmic error distribution  \citep{chen2002semiparametric} in this paper. The survival function of the error $\epsilon$ is given by $\bar{F}_{\epsilon} (t)=e^{-G(e^t)}$. For the logarithmic error distribution class, we have,
\begin{eqnarray}
    G(t)=\frac{log(1+rt)}{r} \hspace{2mm} \textit{if} \hspace{2mm} r>0 \\
    G(t)=t \hspace{2mm} \textit{if} \hspace{2mm} r=0 \notag.
\end{eqnarray}

We get the extreme value and the logistic error distribution for $r=0$ and $r=1$, which leads to the well known PH and PO model respectively. Another possible choice  for the error distribution are the class of Box-Cox type error distributions \citep{zeng2006efficient}, where $G(t)=\frac{(1+t)^\rho-1}{\rho} \hspace{2mm} \textit{if} \hspace{2mm} \rho>0$ and $G(t)=log(1+t) \hspace{2mm} \textit{if} \hspace{2mm} \rho=0$ respectively.

\section{Estimation}
\label{est}
We assume that ${F}_{\epsilon}$ is known and denote the unknown parameters as $\phi=(\bm\beta,\beta(\cdot),H(\cdot))$, with their true values being $\phi_0=(\bm\beta_0,\beta_0(\cdot),H_0(\cdot))$. Further we assume that, $\tau=\inf\{t:P(T\wedge C>t)=0\} <\infty$. We model both the unknown nonparametric functions $H(\cdot)$ and $\beta(s)$ in model (\ref{ttf2}) using univariate Bernstein basis polynomial expansions. The transformation function $H(\cdot)$ is modelled as,
\begin{eqnarray}
    H_{N_0}(t)= H_{N_0}(t;\tau,\bm\gamma)  &=\sum_{k=0}^{N_0}\gamma_{k}b_k(t/\tau,N_0).
     \hspace{1mm} 
    \end{eqnarray}
Here $b_k(x,N)={N \choose k}x^k(1-x)^{N-k}, \hspace{2mm}\hspace{1mm } 0\leq x\leq 1.$. The number of basis polynomials depends on the order of the polynomial basis $N_0$. Note that $b_k(x,N_0)\geq 0$ and $\sum_{k=0}^{N_0}b_k(x,N_0)=1$. The unknown parameters of interest are then the basis coefficients $\bm\gamma=(\gamma_0,\gamma_1,\ldots,\gamma_{N_0})^T$. Based on (3), the derivative of the transformation function $ H_{}^{'}(t)$ is modelled as,
\begin{eqnarray}
     H_{N_0}^{'}(t)&=N_0\sum_{k=0}^{N_0-1}(\gamma_{k+1}-\gamma_{k})b_k(t/\tau,N_0-1) \times (1/\tau).
\end{eqnarray}
Let $ H(\cdot)\in \mathcal{F}$, denote the space of continuous, bounded and increasing transformation functions considered in this paper. We define the constrained Bernstein polynomial sieve following \cite{mclain2013efficient,ghosal2023shape} given by,
\begin{equation}
    \mathcal{F}_{N}=\{H_{N}(t);M_l\leq\gamma_0\leq\gamma_1\leq,\ldots,\leq\gamma_N \leq M_u \}, N=1,2,\ldots\label{sieve}.
\end{equation}
Our estimation procedure enforces the constraint $H_{N}(\cdot)\in\mathcal{F}_{N} \subset \mathcal{F}$. Let $\Gamma_R^{N}$ denote the space of restricted parameters $\bm\gamma$ given by $\Gamma_R^{N}=\{\bm\gamma \in \amsmathbb{R}^{N}  \in ;\gamma_0\leq\gamma_1\leq,\ldots,\leq\gamma_N \}$. It can be easily seen that the sequence of function spaces $\mathcal{F}_{N}$ is nested in $\mathcal{F}$
and $\bigcup_{N=1}^{\infty}\mathcal{F}_{N}$ is dense in $\mathcal{F}$ with respect to the sup-norm \citep{wang2012shape}. This result along with the Bernstein-Weierstrass approximation theorem guarantee that $H_{N}(t)$ converges uniformly to $H_0(t)$ as $N\rightarrow{\infty}$, for $\gamma_k=H_0(K/N)$, for $k=0,\ldots,N$.

The functional coefficient $\beta(s)$ is modelled using Bernstein basis expansion as 
\begin{equation}
    \beta_{N_1}(s)=\sum_{k=0}^{N_1}\theta_{k}b_k(s,N_1).
\end{equation}
Similarly based on Bernstein-Weierstrass approximation theorem, we know that $\beta_{N_1}(s)$ converges uniformly to $\beta(s)$ as $N_1\rightarrow{\infty}$, for $\theta_k=\beta_0(K/N_1)$, for $k=0,\ldots,N_1$.
Let $\*N=(N_0,N_1)$, plugging in these basis expansions, the conditional survival and density functions of $T$ is given by,
\begin{align}
      S_\*N(t|\*X,X_{f}(\cdot),\bm\beta,\bm\gamma,\bm\theta)&=\bar{F}_{\epsilon}\{ H_{N_0}(t)+\bm\beta^T\*X +\sum_{k=0}^{N_1}\theta_k\int_{0}^{1} X_{f}(s)b_k(s,N_1)ds\}\\
     f_\*N(t|\*X,X_{f}(\cdot),\bm\beta,\bm\gamma,\bm\theta)&= {f}_{\epsilon}\{ H_{N_0}(t)+\bm\beta^T\*X +\sum_{k=0}^{N_1}\theta_k\int_{0}^{1} X_{f}(s)b_k(s,N_1)ds\}*H_{N_0}^{'}(t)
\end{align}
The log-likelihood function for this right-censored scenario is given by,
\begin{align}
      l_n(\bm\beta,\bm\gamma,\bm\theta|\+X,\*X_{f}(\cdot),\*Y)&=
      \sum_{i=1}^{n} \Delta_i log f_\*N(Y_i|\*X_i,X_{fi} (\cdot),\bm\beta,\bm\gamma,\bm\theta)+(1-\Delta_i) log S_\*N(Y_i|\*X_i,X_{fi}(\cdot),\bm\beta,\bm\gamma,\bm\theta).
\end{align}
The maximum likelihood estimate of the parameters $\bm\psi=(\bm\beta^T,\bm\gamma^T,\bm\theta^T)^T$ are given by
\begin{equation}
    \hat{\bm\psi}=\underset{\bm\psi \in \Psi_R}{\text{argmax}}\hspace{2 mm} l_n(\bm\psi|\*X,X_{f}(\cdot),\*Y), \Psi_R= \amsmathbb{R}^p\times \Gamma_R\times\amsmathbb{R}^{N_1} .
\end{equation}
Once we have the estimates $\hat{\bm\beta}$,$\hat{\bm\gamma}$ and $\hat{\bm\theta}$, the estimates of the functions $H_0(t)$ and $\beta(s)$ are given by
$\hat{H_0}(t)=\sum_{k=0}^{N_0}\hat{\gamma}_{k}b_k(t/\tau,N_0)$ and $\hat{\beta}(s)=\sum_{k=0}^{N_1}\hat{\theta}_{k}b_k(s,N_1)$, respectively.
Note that the monotonicity restrictions on $\bm\gamma$ given by $\gamma_0\leq\gamma_1\leq,\ldots,\leq\gamma_N$, can be enforced by reparametrizing $\eta_0=\gamma_0$ and $exp(\eta_k)=\gamma_k-\gamma_{k-1}$, for $k=1,\ldots,N$. 

\subsection{Variance Estimation and Inference}
The variance covariance matrix of the maximum likelihood estimates $\hat{\bm\psi}$ are estimated based on the observed Fischer information matrix at the MLE. In particular, the observed Fischer information matrix is given by $\mathcal{J}(\hat{\bm\psi})=-\frac{\partial^2 l_n(\bm\psi|\*X,X_{f}(\cdot),\*Y)}{\partial \bm\psi \partial \bm\psi^T}|_{\bm\psi=\hat{\bm\psi}}$. The estimates of the asymptotic variance covariance matrix of the estimates then are given by $\hat{V}(\hat{\bm\psi})=(\mathcal{J}(\hat{\bm\psi}))^{-1}$. Once the estimates of $\hat{\bm\beta}$, $\hat{H_0}(t)$ and $\hat{\beta}(s)$ are obtained, we define the Wald confidence intervals (point-wise for the functional parameters) as estimate$\pm$ $1.96$* se, based on the estimates and their estimated standard errors. Theoretical details regarding consistency %and 
%asymptotic normality 
of the estimators are provided in Appendix A of the Supplementary Material, under standard regularity conditions, following the results in \cite{mclain2013efficient}. 

\subsection{Choice of Tuning Parameters}
The number of basis functions $N_0$ and $N_1$ are a function of sample size $n$ and are chosen in a data-driven way. Asymptotically, for both of these parameters, we have $N_j=O(n^k_j), k_j\in (0,1)$. The parameter $N_0$ controls the smoothness of the transformation function $H$. The parameter of $N_1$ controls the smoothness of the functional coefficient $\beta(s)$. In this paper, we follow a truncated basis approach, by restricting the number of Bernstein polynomial bases to incorporate smoothness \citep{Ramsay05functionaldata,fan2015functional}. In finite sample size, we use the Akaike information criterion (AIC) defined below to choose $N_0,N_1$ using a a grid-search providing the empirically minimum AIC, 
\begin{equation}
    AIC(\*N)=AIC(N_0,N_1)=-2 l_n(\bm\beta,\bm\gamma,\bm\theta|\*X,X_{f}(\cdot),\*Y)+2(p+N_0+N_1+1).
\end{equation}
%Throughout this article, the optimal choice of $N_0,N_1$ are chosen using a grid-search providing the empirically minimum AIC.

\section{Simulation Study}
 \label{simul}
In this Section, we investigate the performance of the proposed estimation and inference method for the proposed FTTM via simulations. To this end, we consider the following data generating scenarios.
\subsection{Data Generating Scenarios}
\subsection*{Scenario A1: Functional Proportional Hazard Model}
We generate the survival
times following the functional Cox model with the PH structure, 
\begin{align*}
    log \lambda_i(t|\*X_i,X_{fi}(\cdot))=log \lambda_{0}(t)+X_{i1}\beta_1+X_{i2}\beta_2  + \int_{0}^{1}X_{fi}(s)\beta(s)ds, i=1,2,\ldots,n.
\end{align*}
The baseline survival time follows an exponential distribution with rate $e^{\beta_0}$, hence the baseline hazard is taken as $\lambda_{0}(t)=e^{\beta_0}=0.2$. The scalar covariates $X_{i1},X_{i2}$ are independently generated from a Bernoulli (0.5) and a standard Normal distribution.  The functional predictors $X_{fi}(\cdot)$ are independently generated as $X_{fi}(s)=\sum_{k=1}^{10}\psi_{ik}\phi_k(s)$, where $\phi_k(s)$ are orthogonal basis polynomials (of degree $k-1$) and $\psi_{ik}$ are  mean zero and independent Normally distributed scores with variance $\sigma^2_k=4(10-k+1)$. They are observed on a a dense and regular grid of length $m=101$ on $\mathcal{S}=[0,1]$. The scalar and the functional coefficients are taken to be
$\bm\beta=(-0.5,0.4)$, $\beta(s)=cos (\pi s)$ respectively. Censoring times are independently generated from an exponential distribution with mean $20$, which results into a censoring rate of approximately $26\%$. We take $\tau$ to be the nearest integer, larger than the maximum observed time. 
As mentioned in Section 2.1 this is a special case of the FTTM with the extreme value error distribution, specified by $r=0$ in our error model (3)  and $H(t)=log(0.2t)$.
We consider three sample sizes $n\in\{100,300,500\}$ and 100  Monte-
Carlo (M.C) replications from the above data generating scenario.

\subsection*{Scenario A2: Functional Proportional Odds Model}
We generate survival times from the FTTM,
\begin{eqnarray}
  S_T(t|\*X,X_{f}(\cdot))=\bar{F}_{\epsilon}\{ H(t)+X_{i1}\beta_1+X_{i2}\beta_2 +\int_{0}^{1} X_{f}(s)\beta(s)ds\} \notag,
  \end{eqnarray}
where we assume $\epsilon$ follows a logistic distribution with survival function $\bar{F}_{\epsilon}(t)=\frac{1}{1+e^t}$ and $H(t)=log(t^2)$. As mentioned in Section 2.1, this is a special case of the FTTM with the logistic error distribution, specified by $r=1$ in our error model (3) and leads to a PO model, with $logit\{F(t|\*X,X_{f}(\cdot))\}=H(t)+X_{i1}\beta_1+X_{i2}\beta_2 +\int_{0}^{1} X_{f}(s)\beta(s)ds$. The scalar and the functional covariates are generated as in scenario A1. The scalar and the functional coefficients are taken to be
$\bm\beta=(-0.8,1.6)$, $\beta(s)=2sin(\pi s)$ respectively. Censoring times are independently generated from an exponential distribution with mean $5$, which results into a censoring rate of approximately $30\%$. We again consider three sample sizes $n\in\{100,300,500\}$ and 100  Monte-
Carlo (M.C) replications from the above data generating scenario.
\subsection{Simulation Results}
\subsection*{Performance under scenario A1}
\begin{figure}[ht]
\centering
\includegraphics[width=0.7\linewidth , height=0.5\linewidth]{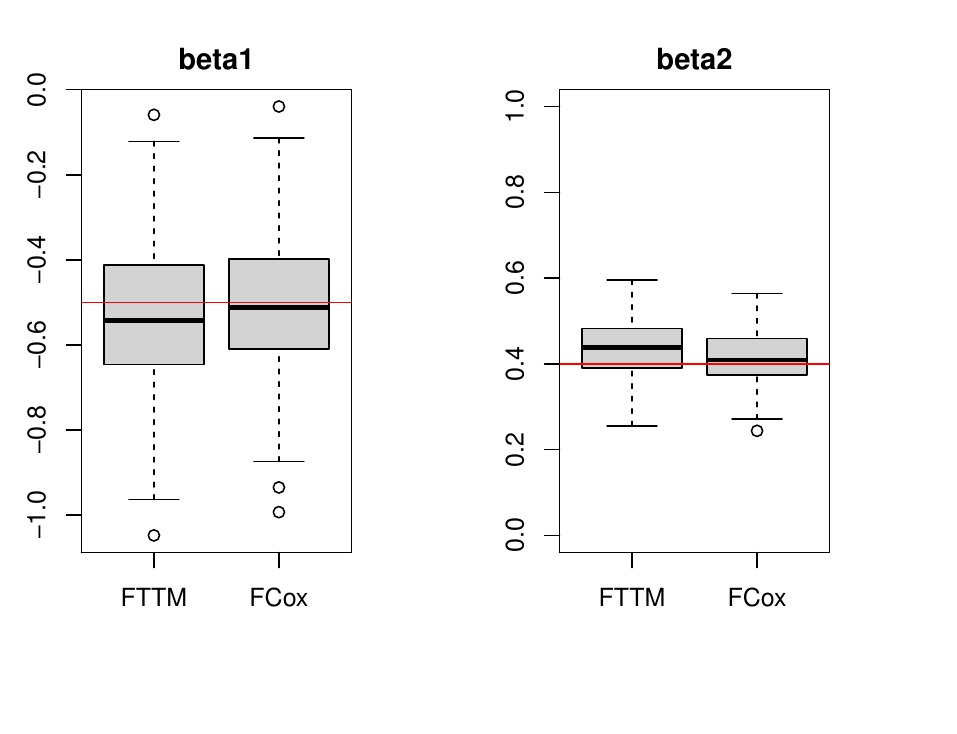}
\caption{Boxplots of $\hat{\beta}_1,\hat{\beta}_2$ from the FTTM and a linear functional Cox model (FCox), scenario A1, n=300. The red solid line indicates the true value of the parameters.}
\label{fig:fig2sc}
\end{figure}

We apply the proposed FTTM to estimate the scalar coefficients $(\beta_1,\beta_2)$, the functional coefficient functions $\beta(s)$ and the transformation function $H(t)$. The extreme value distribution with $r=0$ corresponding to equation (3) is used as the error distribution. The optimal order of BP basis $N_0,N_1$ are chosen using a grid-search 
over $N_0\in \{4,7,10,13\}$ and $N_1\in \{3,5,7,9\}$, providing the minimum AIC value in (12). We also fit the linear functional cox model \citep{gellar2015cox} for comparison purpose, which is the true generating model in this case and avoids smooth semi-parametric estimation of $H(t)$.
The distribution of the estimated scalar parameters from both the approach are displayed in Figure \ref{fig:fig2sc}, for sample size $n=300$. The estimates from the FTTM can be noticed to be slightly more biased compared to the oracle functional Cox model.  

\begin{figure}[ht]
\begin{center}
\begin{tabular}{ll}
\includegraphics[width=.48\linewidth , height=.45\linewidth]{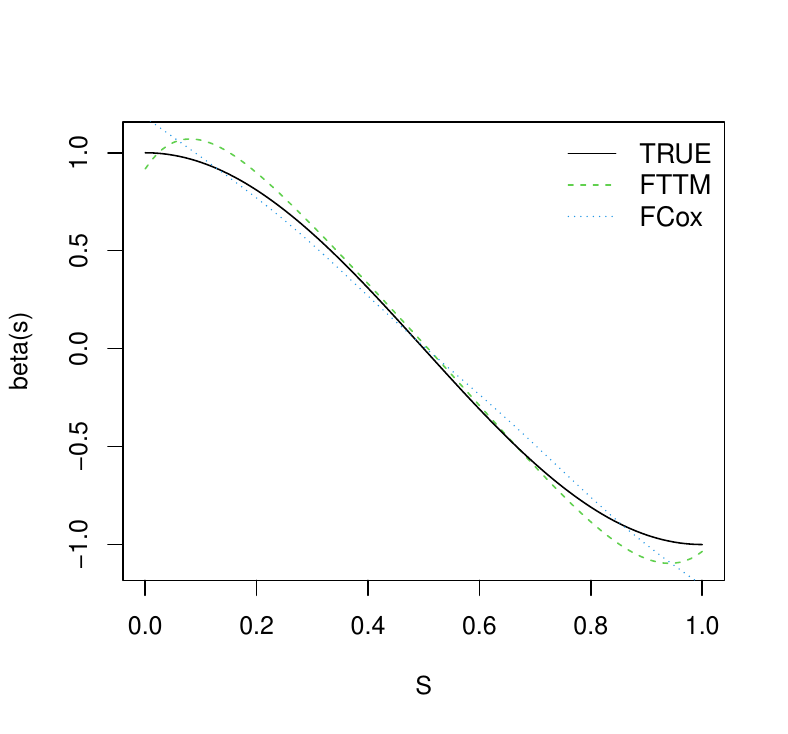} &
\includegraphics[width=.48\linewidth , height=.45\linewidth]{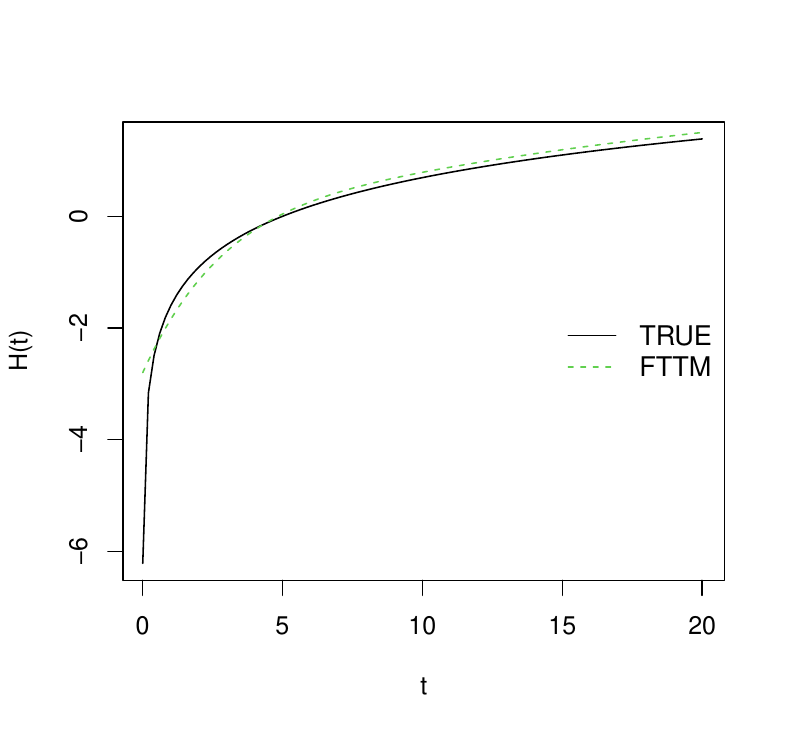}
\end{tabular}
\end{center}
\caption{Left: True functional effect $\beta(s)$ (solid) and estimated $\hat{\beta}(s)$ averaged over 100 M.C replications from the FTTM (dashed) and functional Cox model (dotted), scenario A1, $n=300$. Right: True $H(t)$ (solid) and estimated $\hat{H}(t)$ from the FTTM averaged over 100 M.C replications (dashed).}
\label{fig:fig3s1}
\end{figure}

The estimates of the functional coefficients $\beta(s)$ and the transformation function $H(t)$ from the FTTM averaged over 100 M.C replications are shown in Figure \ref{fig:fig3s1}, for sample size $n=300$. The average estimated coefficient function for $\beta(s)$ from the linear functional cox model is also included for comparison. We observe that the true functional coefficient $\beta(s)$ are estimated accurately by both the FTTM and the functional Cox model, although the FTTM exhibits a slightly higher point-wise bias. The transformation function $H(t)$ is also closely approximated by the M.C mean estimate from the FTTM, except for very small values of $t$, where a positive bias can be noticed.

We report the Monte Carlo (M.C) mean square error (MSE) of the estimated scalar parameters and 
the mean integrated squared error ($MISE=\frac{1}{M} \sum_{j=1}^{M}\int_{0}^{1}\{\hat{\beta}^{j}(s)-\beta(s)\}^2ds$) of the estimates of the functional parameter ($\beta(s)$) for both the FTTM and the functional Cox model in Table \ref{tab:my-table1} across all three sample sizes. For the FTTM we also report the MISE of the estimated transformation function $H(\cdot)$ obtained over a grid of time-points.

\begin{table}[H]
\centering
\caption{Monte Carlo mean squared error of the estimated scalar and functional parameters from the functional time transformation model (FTTM) and the oracle functional Cox model, under Scenario A1. The Cox model performance is given in the parenthesis.}
\label{tab:my-table1}
\begin{tabular}{lllll}
\hline
Sample Size & MSE $\beta_1$   & MSE $\beta_2$   & MISE $\beta(s)$ & MISE $H(t)$ \\ \hline
n=100       & 0.110 (0.082) & 0.025 (0.0171) & 0.448 (0.142) & 0.211   \\ \hline
n=300       & 0.030 (0.025) & 0.006 (0.004) & 0.124 (0.045) & 0.149 \\ \hline
n=500       & 0.016 (0.013) & 0.003 (0.003) & 0.067 (0.030) & 0.152 \\ \hline
%n=1000      & 0.0097 (0.0069) & 0.0024 (0.0013) & 0.037 (0.017)  & -0.046 (-0.014) & 0.031 (0.007)  \\ \hline
\end{tabular}
\end{table}

The proposed estimation method for FTTM does a satisfactory job in estimating the true scalar and functional parameters, with accuracy improving with increasing sample size. The functional Cox model is the oracle model in this scenario, which avoids estimation of $H$ (an additional parameter estimated in our FTTM) by using partial log-likelihood \citep{gellar2014variable}, and produces a better performance which is expected, on account of being less biased compared to the FTTM in this case.

We also explore the coverage of the Wald confidence intervals developed in Section 2.3 based on the model based standard errors, for the scalar and functional coefficients. We set the the tuning parameter combination to be $N_0=13,N_1=3$ for this analysis, which was the best performing parameters (providing the lowest AIC) across the 100 M.C replications while estimation. The estimated coverage from $95\%$ Wald confidence intervals across 100 Monte Carlo replications are presented in Supplementary Table S1. We report the average point-wise coverage over $S$ for the functional coefficient $\beta(s)$. The estimated coverage can be observed to be close to the nominal level of $0.95$ for both the scalar and functional parameters indicating a satisfactory performance of the proposed method.

\subsection*{Performance under scenario A2}
\begin{figure}[H]
\centering
\includegraphics[width=0.7\linewidth , height=0.5\linewidth]{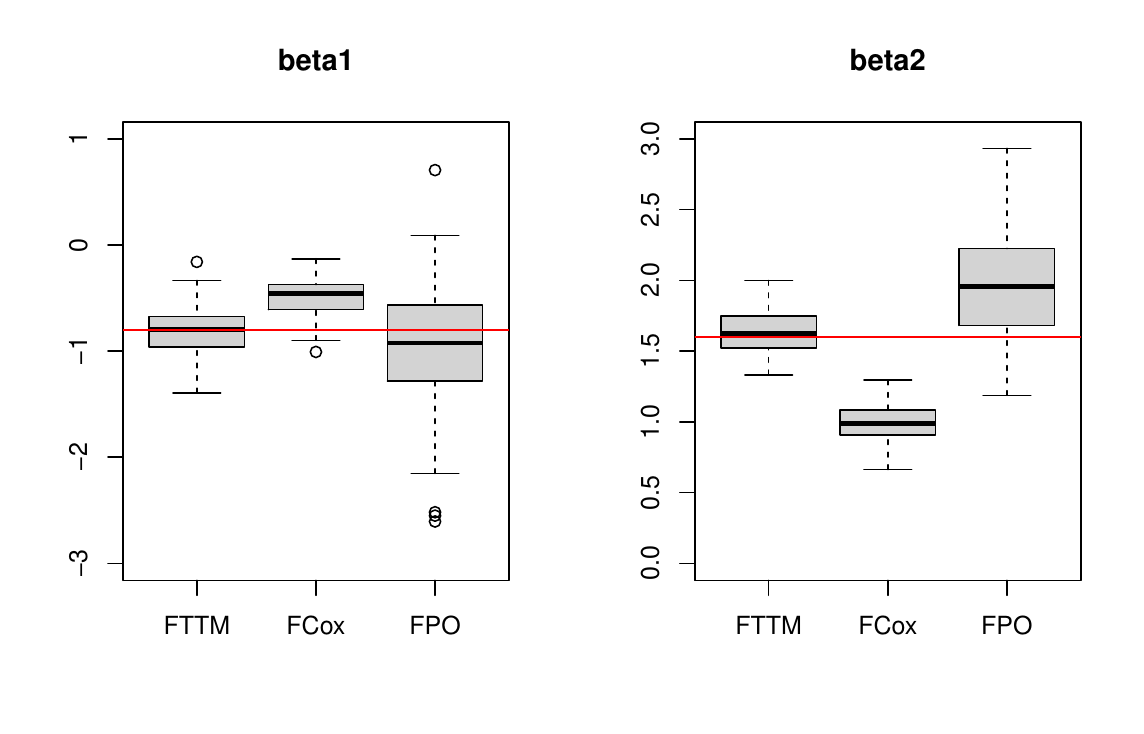}
\caption{Boxplots of $\hat{\beta}_1,\hat{\beta}_2$ from the FTTM, a linear functional Cox model (FCox), and the FPO model under scenario A2, n=300. The red solid line indicates the true value of the parameters.}
\label{fig:fig4sc}
\end{figure}

We again apply the proposed FTTM to estimate the model parameters. The error distribution used in this case is the logistic distribution with $r=1$ corresponding to equation (3). The optimal order of BP basis $N_0,N_1$ are chosen using a grid-search providing the minimum AIC value as in the previous scenario. As competing models, We fit i) the linear functional cox model (termed as FCox), which will be misspecified in this scenario, and ii) a functional extension of the semiparametric proportional odds model proposed in \cite{eriksson2015proportional}. In this approach (termed as FPO), we use the functional principal component (FPC) scores of the functional covariate $X(s)$ explaining more than $95\%$ variability, as additional scalar predictors. The distribution of the estimated scalar parameters from both the approach are displayed in Figure \ref{fig:fig4sc}, for sample size $n=300$. The estimates from the FTTM can be noticed to accurately capture the true parameters, where as both FCox and FPO lead to high biases.  

\begin{figure}[ht]
\begin{center}
\begin{tabular}{ll}
\includegraphics[width=.48\linewidth , height=.45\linewidth]{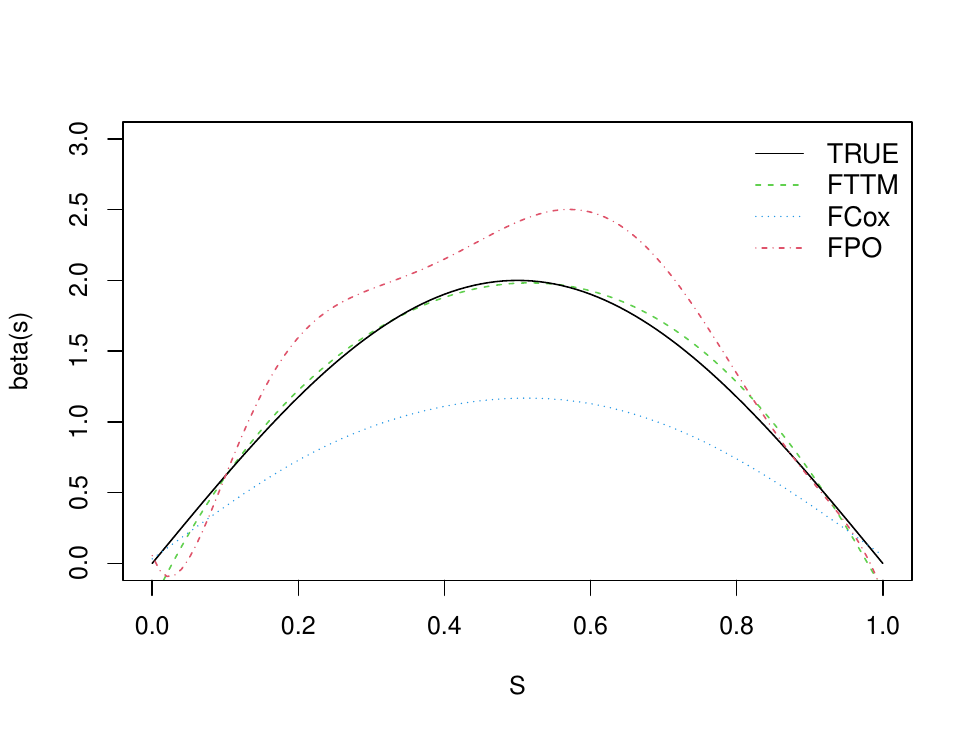} &
\includegraphics[width=.48\linewidth , height=.45\linewidth]{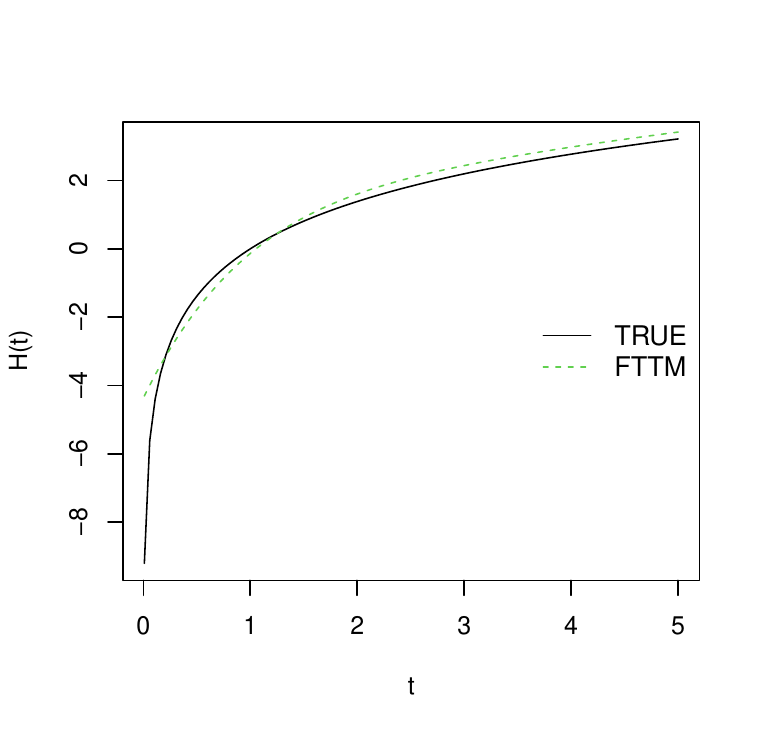}
\end{tabular}
\end{center}
\caption{Left: True functional effect $\beta(s)$ (solid) and estimated $\hat{\beta}(s)$ averaged over 100 M.C replications from the FTTM (dashed), FCox (dotted), FPO (dashed-dotted), under scenario A2, $n=300$. Right: True $H(t)$ (solid) and estimated $\hat{H}(t)$ from the FTTM averaged over 100 M.C replications (dashed).}
\label{fig:fig5s1}
\end{figure}

The M.C mean estimates of the functional coefficients $\beta(s)$ and the transformation function $H(t)$ from the FTTM are shown in Figure \ref{fig:fig5s1}, for sample size $n=300$. The average estimated coefficient function for $\beta(s)$ from the FCox and FPO is also included for comparison. We observe that the true functional coefficient $\beta(s)$ are estimated accurately by the FTTM, whereas the FCox and FPO estimate exhibit a high bias. Moreover, the FPO estimate is highly non-smooth, whereas the FTTM produces a smooth estimate of the functional coefficient due to the truncated basis selection. The transformation function $H(t)$ is again closely estimated from the FTTM, except for very small values of $t$.

The Monte Carlo (M.C) mean square error (MSE) of the estimated scalar and functional parameters for the FTTM, FCox, and the FPO model are reported in in Table \ref{tab:my-table2} across all three sample sizes. The MISE of the estimated transformation function $H(\cdot)$ is also reported for FTTM. A satisfactory performance of FTTM can be observed for all the parameters especially for larger sample sizes.

\begin{table}[H]
\centering
\caption{Monte Carlo mean squared error of the estimated scalar and functional parameters from the functional time transformation model (FTTM), FCox and FPO model under scenario A2. The FCox and FPO model performances are given in the parenthesis.}
\label{tab:my-table2}
\begin{tabular}{lllll}
\hline
Sample Size & MSE $\beta_1$   & MSE $\beta_2$   & MISE $\beta(s)$ & MISE $H(t)$ \\ \hline
n=100       & 0.251 (0.188,0.889) & 0.114 (0.320,0.506) & 0.819 (0.554,9.360) & 0.568   \\ \hline
n=300       & 0.049 (0.125,0.415) & 0.025 (0.383,0.280) & 0.212 (0.409,3.87) & 0.375 \\ \hline
n=500       & 0.028 (0.109,0.383) & 0.012 (0.421,0.208) & 0.111 (0.379,2.92)  & 0.378 \\ \hline
%n=1000      & 0.0097 (0.0069) & 0.0024 (0.0013) & 0.037 (0.017)  & -0.046 (-0.014) & 0.031 (0.007)  \\ \hline
\end{tabular}
\end{table}

Overall, the proposed estimation method for FTTM clearly outperforms the FCox model, on average producing $2.4,17.7$ times smaller MSE for the scalar parameters $\beta_1,\beta_2$, and 2 times smaller MISE for the functional parameter $\beta(s)$. Similarly, compared to the FPO model, the FTTM on average produces $8.6,11$ times smaller MSE for the scalar parameters $\beta_1,\beta_2$, and $18.7$ times smaller MISE for the functional parameter $\beta(s)$. The improvement in the performance is particularly pronounced for the higher sample sizes.
As in scenario A1, we report the estimated coverage of the parameters from $95\%$ Wald confidence intervals across 100 Monte Carlo replications in Supplementary Table S2,for the tuning parameter choice $N_0=13,N_1=3$. The estimated coverage is observed to be close to the nominal level of $0.95$ for both the scalar and functional parameters.

%Additional simulation scenarios are explored in Appendix E of the supplementary material where model selection performance between different parametric models are investigated using the proposed AIC approach.

\section{Real Data Application}
\label{realdat}
\subsection{Modelling All-cause Mortality in NHANES 2011-2014}
In this study, we are interested in quantifying the association between diurnal patterns of objectively measured physical activity and all-cause mortality among the older adult population (aged 50 years or older) in USA. We apply the proposed FTTM to physical activity data collected via accelerometer from the National Health and Nutrition Examination Survey (NHANES) 2011-2014 and all-cause mortality data through 2019 while adjusting for age, gender and body mass index (BMI) at baseline. The NHANES provides a broad range of descriptive health and nutrition statistics and is a nationally representative sample of the non-institutionalized US population.  In the NHANES 2011-2014, accelerometry data was collected using the ActiGraph GT3X+ accelerometer worn on the wrist (manufactured by ActiGraph of Pensacola, FL). Participants were instructed to wear the activity monitor continuously for seven days, removing it on the morning of the 9th day. Our analysis focuses on the minute-level accelerometer data from 2011 to 2014, which was made available in 2021. This data reports individuals' acceleration in Monitor Independent Movement Summary (MIMS) units, an open source device-independent metric for summarizing movement activity \citep{john2019open}. It is possible to link NHANES 2011-2014 data to the National Death Index (NDI) \citep{leroux2019organizing} for collecting mortality information. For this purpose, we use the latest (December 31, 2019) mortality information from NDI (\url{https://www.cdc.gov/nchs/data-linkage/mortality-public.htm}) to define our survival outcome.

A total of 3032 adults aged 50 years or older (with physical activity monitoring available at least ten hours per day for at least four days) with available mortality and covariate (age, gender, and BMI) information were included in our analysis. The descriptive statistics of the sample is reported in supplementary Table S3. We denote the log-transformed MIMS as $X_{ij}(s)$, $s\in [0,1440]$, which represent the log-transformed MIMS ( $A\rightarrow log(1+A)$) for the individual $i$ on the day $j$ at time $s$, $i=1,\dots,n$, and $j=1,\dots,n_i$. The diurnal average functional curve  $X_i(s)= \frac{1}{n_i} \sum_{j=1}^{n_i} X_{ij}(s)$  is considered as the functional covariate of interest. The diurnal curves $X_i(s)$ are further aggregated into 10 minutes epochs to make the PA profiles smoother for each subject. Survival time is calculated in years from the end of accelerometer wear, with all subjects being censored on December 31, 2019. Among the 3032 study participants considered at the baseline, 582 ($19.2\%$) were deceased by the end of the study. The mean follow-up time of the subjects is 6.4 years. Figure \ref{fig:fig6s1} left panel displays the average diurnal PA profile of subjects for the deceased group and the survivors. The deceased group can be observed to have lower PA compared to the survivors particularly during the day-time.

\subsubsection*{FTTM for modelling all-cause mortality}
We fit the following FTTM developed in this article, to directly model all-cause survival based on the diurnal PA profile and age, gender ($G_i=1$ for Female) and BMI.

\begin{equation}
      H(T_i)
    = -\{\beta_1Age_i +\beta_2G_i +\beta_3BMI_i +\int_{0}^{1} X_{i}(s)\beta(s)ds\}+\epsilon_i. 
\end{equation}

We assume $\epsilon$ is coming from the class of logarithmic error distribution described in error model (3), where the distribution function $F_{\epsilon}(\cdot)$ is known for a fixed value of $r$. Instead of fixing the value of $r$, we try several candidate models varying $r\in\{0,0.05,0.1,\ldots,0.1\}$, and chose the one which produces the minimum AIC proposed in Section 2.4. Note that this includes the Cox model (PH) and the PO model as special cases for $r=0,1$ respectively. The tuning parameters $N_0\in\{6,9,12,15,18\}$ and $N_1\in\{2,3,4,5\}$ are also varied to control the smoothness of the transformation function and the functional coefficient. The optimal combination of $N_0,N_1,r$ chosen using the proposed AIC is given by $N_0=9, N_1=4, r=0.35$ producing an AIC of $4715.1$. 

\begin{figure}[H]
\begin{center}
\begin{tabular}{ll}
\includegraphics[width=.5\linewidth , height=.5\linewidth]{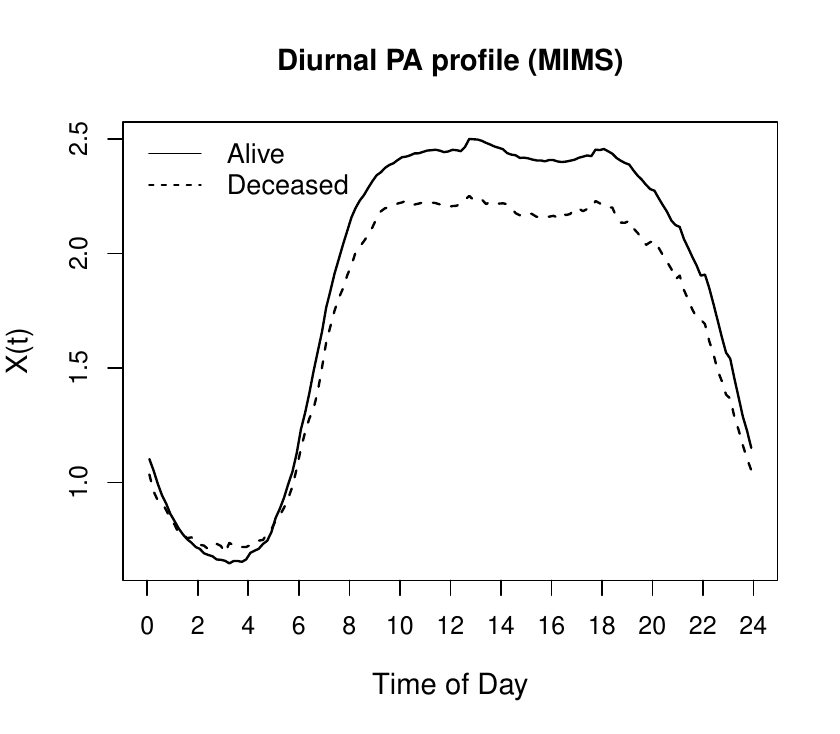} &
\includegraphics[width=.5\linewidth , height=.5\linewidth]{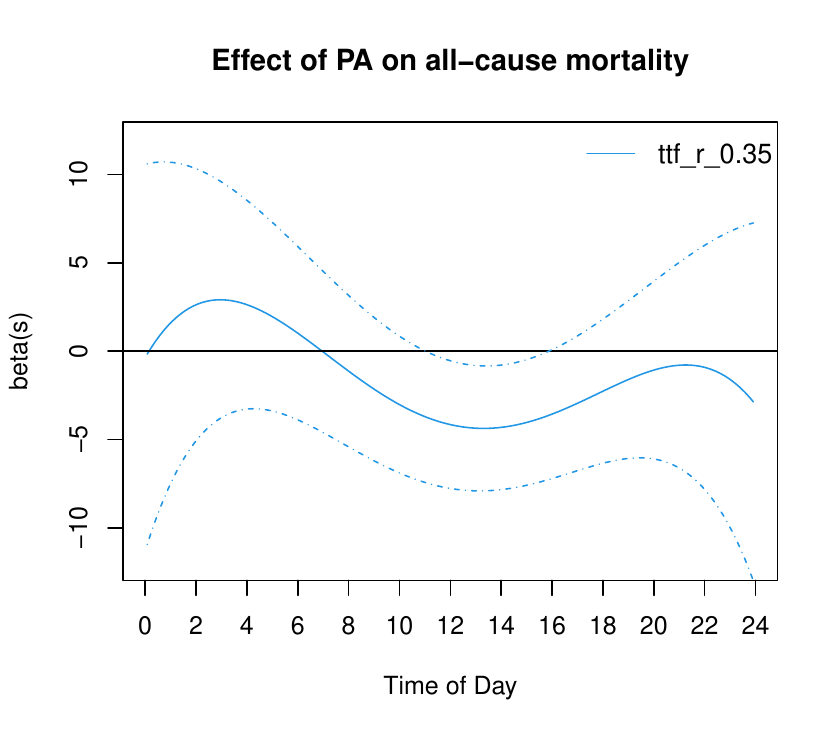}
\end{tabular}
\end{center}
\caption{Left: Average diurnal PA profile (log-MIMS) of subjects for the deceased group (dashed) and the survivors (solid) in the NHANES application. Right: Estimated dynamic effect $\hat{\beta}(s)$ (solid) from the FTTM capturing the effect of diurnal PA on all-cause mortality and it's associated $95\%$ point-wise confidence interval (dashed and dotted).}
\label{fig:fig6s1}
\end{figure}
The estimated functional coefficient $\hat{\beta}(s)$ along with the corresponding $95\%$ point-wise confidence interval is displayed in Figure \ref{fig:fig6s1} right panel. The estimated effect $\hat{\beta}(s)$ is negative during the daytime 11 am - 4pm, indicating a higher diurnal PA during these times of day is associated with a higher survival time (or it's increasing transformation) and a higher survival probability following model (13). The estimated scalar coefficients and their associated $95\%$ confidence intervals are reported in Supplementary Table S4, where we observe a higher age to be associated ($\hat{\beta}_1=0.8$) with a reduced survival probability.

We display the predicted survival probability plots $\hat{S}_T(t|\*X,X(\cdot))=\bar{F}_{\epsilon,r=0.35}\{ \hat{H}(t)+\hat{\beta}_1Age_i +\hat{\beta}_2G_i +\hat{\beta}_3BMI_i +\int_{0}^{1} X_{i}(s)\hat{\beta}(s)ds\} $ in Supplementary Figure S2 for male and female subjects with age = 66 years at baseline and having BMI= 30.06 $Kg/m^2$, using the mean value of age, BMI at baseline and the diurnal PA curve $X(s)$ belonging to either of high, average, or low group. The “High”, “average” and “low” group was defined as having $X (s)=\bar{X}(s)+0.5$, $X (s)=\bar{X}(s)$, and $X (s)=\bar{X}(s)-0.5$ at baseline respectively for illustrative purposes, where $\bar{X}(s)$ denotes the sample average. We notice that higher PA leads to a better survival for both the genders, illustrating a protective effect of PA.

\subsubsection*{Graphical check of model fit}
Note that since the optimal $r$ (0.35) chosen by the AIC is non-zero, this indicates a departure from the PH model. To validate our model, we propose a goodness-of-fit check based on the fitted survival function, which is similar to a graphical check using Cox-Snell residuals. If the model is correctly specified, $-log\{\hat{S}_T(T_i|\*X_i,X_i(\cdot))\}\sim exp(1)$, hence the observations $U_i=-log\{\hat{S}_T(Y_i|\*X_i,X_i(\cdot))\}$ should behave like a censored sample from an exponential distribution with mean 1. This can be graphically checked by using the Nelson-Allen estimator based on fitted $U_i$'s, $i=1,\ldots,n$, which we term as FTTM pseudo residuals. If the fit is good, we can expect $\hat{\Lambda}(U_i)=U_i$, for an unit mean exponential distribution. The estimated cumulative hazard function of $U_i$s are shown in Supplementary Figure S2, which closely follows the 45 degree line ${\Lambda}(U_i)=U_i$. Hence, a satisfactory model fit can be concluded based on the FTTM analysis.

\subsection{Modelling Time to Multiple Hypoglycemia events in CGM Study}
In this application, the high dimensional physiological signal we consider is continuous glucose monitoring (CGM) data collected by the Juvenile Diabetes Research Foundation (JDRF) CGM study group \citep{juvenile2009effect} in a cohort of Type 1 diabetes mellitus (T1D) patients. This was the first landmark randomized clinical trial (RCT) investigating the efficacy and safety of CGM. Continuous Glucose Monitoring (CGM) plays an important role in
diabetes management and can help in improving glycemic control \citep{burge2008continuous,allen2009continuous}. The adverse event, we consider in this study is Hypoglycemia \citep{van2016continuous,hermanns2019impact}, which arises when blood sugar (glucose) level is lower than the standard range. Although prior research over the last decade has demonstrated that real-time continuous glucose monitoring improves glycemic control in terms of lowering glycated hemoglobin levels (HbA1c), such significant effect has not been established for mild or severe hypoglycemia \citep{van2016continuous}.  

Traditionally CGM data have been analyzed using summary level metrics such as the mean \citep{battelino2023continuous},
which can result in substantial loss of information. As an example, consider two individuals
with similar mean CGM, but with very different profiles, one stable around the mean and another with long excursions into the hyper and hypo-glycemic ranges. These two individuals would require different interventions, but
would be hard to distinguish based on mean CGM or HbA1c (average blood sugar (glucose) over past three months), the primary biomarker used for diabetes. Hence, in this article, we propose to use distributional representation of CGM data at baseline (spanning a 10 day period) via subject-specific quantile function $Q_i(p)$ \citep{ghosal2021distributional}
as a functional predictor which is directly related to glucodensities introduced in \cite{matabuena2021glucodensities}. Specifically, We focus on the empirical quantile function of each participant's log-transformed glucose measurements in this application. This is defined as $\hat{Q}_i(p)$, where $p\in [0,1]$ and this can be obtained as the generalized inverse of the empirical cumulative distribution function (CDF) associated with the participant's glucose levels. The empirical CDF, $\hat{F}_{i}(a)$, captures the proportion of glucose measurements that do not exceed a certain level $a$, is given by $\hat{F}_{i}(a) = \frac{1}{n_i} \sum_{j=1}^{n_i} \mathbf{1}\{G_{ij} \leq a\}$, where $G_{ij}, j=1,2,\ldots n_i$ are the glucose values recorded for the $i$-th participant.

In our analysis, we consider a subsample of $296$ subjects from the JDRF study, with available age, Gender, baseline HbA1c and baseline CGM information. The descriptive statistics of the sample is presented in Supplementary Table S5. We define episodes of hypoglycemia as a plasma glucose of $\leq 70$ mg/dl. We define our survival outcome as time to 65 hypoglycemic events over a maximum follow up period of $240$ days which correspond to a subject having mild \citep{ostenson2014self} hypoglycemia (1-2 events per patient per week). In our sample of $296$ subjects, 260 ($87.8\%$) subjects had the adverse event (65 hypoglycemic episodes) by the end of the followup, with the mean follow-up time being 51.2 days. Figure \ref{fig:fig7s1} left panel displays the average quantile function (Wasserstein Barrycenter) $Q_G(p), G=1,2$ of log-transformed glucose values for two groups, subjects who experienced the adverse event (``Events") and those who didn't (``No events"). The events group can be observed to have a lower minimal and maximal quantiles compared to the no events group. We are interested in directly modelling the survival time of this adverse event based on the glucodensity representation $Q_i(p)$ while adjusting for age, Gender, and also investigate whether glucodensity can provide an improved predictive capacity for assessing the risk of hypoglycemia compared to traditional glucose biomarkers such as hemoglobin A1c (HbA1c).  

\subsubsection*{FTTM for modelling time to multiple Hypoglycemia}
We fit the FTTM developed in this article, to directly model the survival time $T_i$ of this adverse event based on the glucodensity representation $Q_i(p)$ and adjusting for age, Gender (($G_i=1$ for Male)).
\begin{equation}
      H(T_i)
    = -\{\beta_1Age_i +\beta_2G_i  +\int_{0}^{1} Q_{i}(p)\beta(p)dp\}+\epsilon_i. 
\end{equation}
The error $\epsilon$ is again assumed to becoming from the class of logarithmic error distribution described in error model (3), where the distribution function $F_{\epsilon}(\cdot)$ is known for a fixed value of $r$. We try several candidate models varying $r\in\{0,0.2,0.4,\ldots,4\}$, and chose the one which produces the minimum AIC. The tuning parameters $N_0\in\{6,9,12,15,18,21\}$ and $N_1\in\{2,3,4,5,6\}$ are also varied to control the smoothness of the transformation function and the functional coefficient. The optimal combination of $N_0,N_1,r$ chosen using the proposed AIC is given by $N_0=21, N_1=4, r=4$ producing an AIC of $2272.3$. In comparison the optimal Cox model ($r=0$) had AIC $2359.4$ and the optimal PO model ($r=1$) had AIC $2314.8$. A high $r=4$ value in this case indicates the covariate effects to be decreasing over time \citep{zeng2006efficient}.

\begin{figure}[ht]
\begin{center}
\begin{tabular}{ll}
\includegraphics[width=.45\linewidth , height=.45\linewidth]{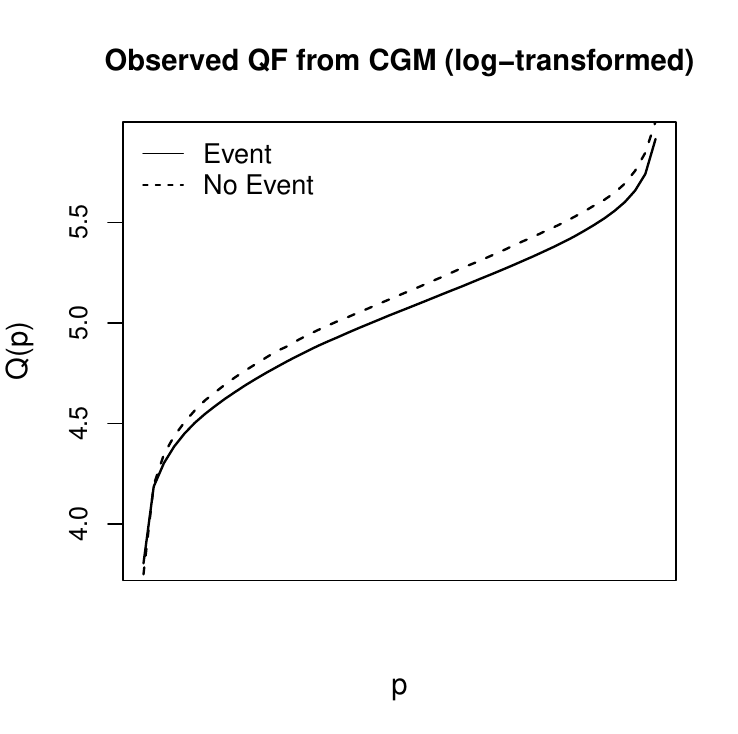} &
\includegraphics[width=.45\linewidth , height=.45\linewidth]{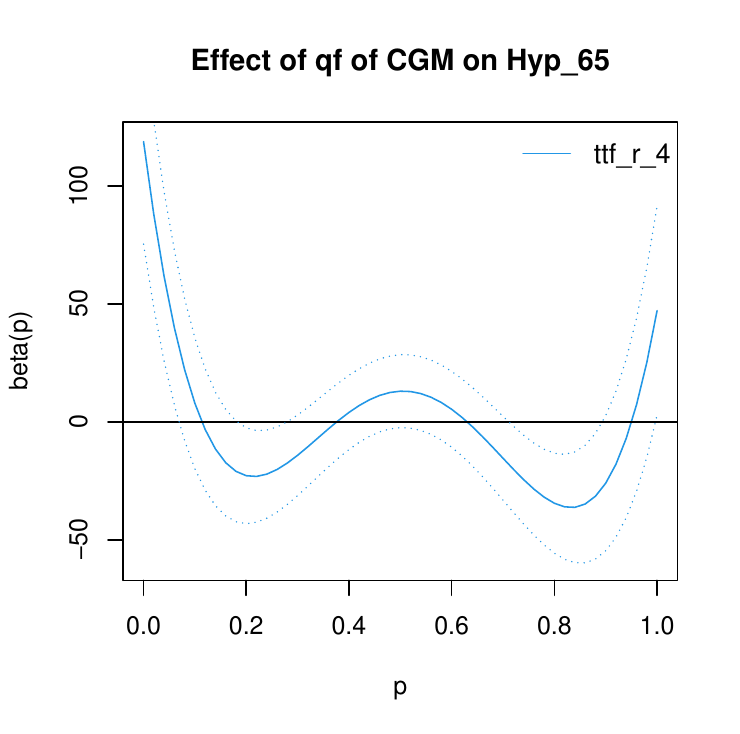}
\end{tabular}
\end{center}
\caption{Left: Average quantile function of log-transformed CGM for the ``Events" group" (solid) and ``No events" group (dashed) in the CGM application. Right: Estimated dynamic effect $\hat{\beta}(p)$ (solid) from the FTTM capturing the distributional effect of CGM on time to 65 Hypoglycemia events and it's associated $95\%$ point-wise confidence interval (dashed and dotted).}
\label{fig:fig7s1}
\end{figure}

The estimated coefficient $\hat{\beta}(p)$ along with the corresponding $95\%$ point-wise confidence interval is displayed in Figure \ref{fig:fig7s1} right panel. The estimated effect $\hat{\beta}(p)$ is found to capture a contrast between CGM values at higher quantile range $p\in[0.7,0.9]$ and CGM values at very low quantile range $p\in[0,0.1]$. A high negative value of this contrast (indicating a higher maximal CGM levels and smaller minimal CGM levels) is found to be associated with a higher survival probability for time to 65 Hypoglycemia. We report the estimated scalar coefficients and their associated $95\%$ confidence intervals in Supplementary Table S6, which were not found to be significant after adjusting for the distributional CGM profile.

\subsubsection*{Graphical check of model fit}
We again use a graphical check based on the fitted survival function, using the FTTM  pseudo residuals $U_i=-log\{\hat{S}_T(Y_i|\*X_i,Q_i(\cdot))\}$ for $r=4$. The estimated cumulative hazard function and the associated $95\%$ confidence interval are shown in Supplementary Figure S3, which are close to the 45 degree line, ${\Lambda}(U_i)=U_i$. Hence, a satisfactory model fit can be concluded based on our analysis.

\subsection*{Predictive comparison with traditional glucose biomarkers}
We compare the predictive performance of the FTTM (14) for $r=4$ using CGM based glucodensity $Q_i(p)$ with i) a Cox model using traditional glucose biomarker HbA1c, while adjusting for age and gender and ii) A functional cox model \citep{gellar2015cox} employing penalized partial likelihood (PPL) based estimation, with $Q_i(p)$ as the functional predictor, and adjusting for age and gender. We obtain the average 10-fold cross-validated Harrell's C-index (concordance index) \citep{harrell1996multivariable} for all the models. The cross-validated C-index is calculated to be $0.641$ for the proposed FTTM, $0.56$ with HbA1c based Cox model and $0.626$ for the distributional Cox model with $Q_i(p)$. This illustrate, with glucodensity based FTTM we get an $14.5\%$ improvement in predictive capacity of the model, compared to the traditional biomarker HbA1c used in a Cox model. The proposed FTTM approach, also exhibit a $2.4\%$ improvement over the functional Cox model with $Q_i(p)$. 

Our results in this section highlight the usefulness of the proposed FTTM in survival analysis with infinite-dimensional functional covariates.

\section{Discussion}
\label{disc}
In this paper we have developed a functional
time-transformation model (FTTM) for estimating the conditional survival function in the presence of both functional and scalar covariates. This provides a flexible way to directly model a monotone transformation of survival time in clinical and epidemiological studies based on high dimensional physiological signals which can be treated as functional covariates. We have used Bernstein polynomials to model the monotone transformation and the smooth functional coefficients, and have employed a sieve method of maximum likelihood for estimation. Since we perform maximum likelihood estimation, we are able to obtain asymptotic standard errors and the point-wise confidence intervals of the model coefficients using the observed information matrix under standard regularity conditions. Theoretical results regarding consistency 
%and asymptotic normality 
of the estimators are also established.

Numerical analysis using simulations
illustrate the satisfactory finite-sample performance of the proposed estimation method for FTTM in accurately estimating the scalar, functional coefficients and the monotone transformation function. In the NHANES 2011-2014 application, the proposed method demonstrates that a higher diurnal PA among older adults during the daytime (11 am-4 pm) is associated with an increased survival probability, even after adjusting for age, complimenting the findings in NHANES 2003-06 \citep{liu2016leisure,cui2021additive}. In the CGM application, the FTTM demonstrates an increased predictive capacity of baseline CGM estimated glucodensity marker compared to the traditional HbA1c biomarker in predicting time to multiple hypoglycemic events in patients with Type 1 diabetes mellitus and also captures a significant association between high maximal quantiles of CGM and lower risk of repeated Hypoglycemia.

There are multiple research directions which remain be explored based on our current research. It would be interesting to develop a formal test of goodness of fit \citep{fernandez2019maximum,chernozhukov2021distributional} for the functional time transformation model based on the pseudo residuals introduced in this paper. Nonparametric tests for independence under censored or missing mechanism could also be adopted for this purpose \citep{matabuena2022kernel}. Another interesting direction of research might be to extend FTTM to accommodate longitudinal or multilevel functional predictors \citep{lin2021functional,cui2022fast}, which are becoming increasingly common with the emergence of intensive longitudinal studies in mHealth. Extending the proposed FTTM to more general classes of survival models \citep{gasperoni2020non,li2022joint} remains plausible and would be areas of future interest. 
%\vspace{- 6mm}
\section*{Supplementary Material}
Appendix A, Supplementary Tables S1-S6 and Supplementary Figures S1-S3 are available online with this article.

\section*{Software}
R software illustration of the proposed method is available with this article and will be made available on Github.

\bibliographystyle{Chicago}
\bibliography{refs}
\end{document}

\def\spacingset#1{\renewcommand{\baselinestretch}%
{#1}\small\normalsize} \spacingset{1}

%%%%%%%%%%%%%%%%%%%%%%%%%%%%%%%%%%%%%%%%%%%%%%%%%%%%%%%%%%%%%%%%%%%%%%%%%%%%%%

\if0\blind
{
  \title{\bf Supplementary Material for Functional Time Transformation Models with Applications to Wearable Data}
\author{Rahul Ghosal$^1$,  Marcos Matabuena$^{2}$, Sujit K. Ghosh$^{3}$\\
     $^{1}$ Department of Epidemiology and Biostatistics, University of South Carolina \\
     $^{2}$Department of Biostatistics, Harvard University, Boston, MA 02115, USA\\
$^{3}$ Department of Statistics, North Carolina State University\\
}
  \maketitle
} \fi

\if1\blind
{
  \bigskip
  \bigskip
  \bigskip
  \begin{center}
    {\LARGE\bf Title}
\end{center}
  \medskip
} \fi

\bigskip

\vfill

\newpage
\spacingset{1.5} % DON'T change the spacing!

\section{Appendix A: Consistency of the Estimators}

We assume the following regularity conditions in order to establish consistency of the estimators:

\begin{enumerate}
\item[I.] The true transformation function $H_{0}(\cdot)$ is increasing and bounded over $(0, \tau]$.
\item[II]. Conditional on $\*X,X_f(\cdot)$ the survival time $T$ is independent of $C$, and \\$S_{C}(\cdot \mid \*X,X_f(\cdot))$ is ancillary for $(\bm\beta, \beta(\cdot),H)$.

\item[III.] For $r=1,2$ the $r$ th derivative of $H_{0}(\cdot)$ exists, is bounded and continuous on $(0, \tau]$, and $\bm\beta$ lies in the interior of a compact set $B \subseteq \mathbb{R}^{p}$.

\item[IV.] The coefficient function $\beta(\cdot)\in \mathcal{K} \subset \mathcal{L}^2[0,1]$, where $\mathcal{K}$ is compact and for $r=1,2$ the $r$ th derivative of $\beta_{0}(\cdot)$ exists and is in $\mathcal{L}^2[0,1]$.

\item[V.] $\*X$ is bounded. $\exists$ $M_1>0$ such that $\operatorname{Pr}\left(\|\*X\| \leq M_1\right)=1$, where $\|\cdot\|$ denotes the Euclidean norm in $\mathbb{R}^{p}$.

\item[VI.] $\exists$ $M_2>0$ s.t  $\Pr(\sup_{s\in [0, 1]} |X_f(s)|\leq M_2)=1$.

%$|X_f(s)|\leq M_2$ for $s\in \mathcal{S}$.

\item[VII.] The c.d.f. of the error distribution $F_{\epsilon}$ is defined on $\mathbb{R}$ and has first, and second derivatives that are continuous and bounded. Furthermore, $f_{\epsilon}(t)$ and $\bar{F}_{\epsilon}(t)$ are log-concave in $t$.

\item[VIII.] The distribution function of the censoring values $F_{C}(\cdot \mid \*X,X_f(\cdot) )$ has first and second derivatives that are continuous and bounded on $(0, \tau]$. Furthermore, the bounds do not depend on $\*X,X_f(\cdot)$.

\item[IX.] Let $N_n=\underset{l\in\{0,1\}}{max}N_{\ell}$, and we have $N_n=O\left(n^{\kappa}\right)$, where $0<\kappa<1$.

\end{enumerate}

Define the metric $d$ as,
\begin{equation*}
d\left\{\left(\bm\beta_{1}, \beta_1(\cdot),H_{1}\right),\left(\bm\beta_{2},\beta_2(\cdot), H_{2}\right)\right\}=\left\|\bm\beta_{1}-\bm\beta_{2}\right\|+\left\|\beta_{1}(\cdot)-\beta_{2}(\cdot)\right\|_{\mathcal{H}
}+\left\|H_{1}-H_{2}\right\|_{F_{Y}}, 
\end{equation*}
where $\left\|H_{1}-H_{2}\right\|_{F_{Y}}=\int\left\{H_{1}(t)-H_{2}(t)\right\}^{2} d F_{Y}(t)$, with $F_{Y}(t)=\operatorname{Pr}(T \wedge C \leq t)$ and $\left\|\beta_{1}(\cdot)-\beta_{2}(\cdot)\right\|_{\mathcal{H}
}=(\int_0^1\left\{\beta_{1}(s)-\beta_{2}(s)\right\}^{2} ds)^{1/2}$.

\begin{theorem}
    Suppose that assumptions (I)-(IX) hold. Then
    $$
d\left\{\left(\hat{\bm\beta}_{n, N_0,N_1},\hat{\beta}(\cdot)_{n, N_1}, \hat{H}_{N_0}\right),\left(\bm\beta_{0},\beta_0(\cdot), H_{0}\right)\right\} \rightarrow 0
$$
\end{theorem}

almost surely, as $n \rightarrow \infty$.

\hspace{- 6 mm}
\textbf{Proof:}
The proof follows the steps of Theorem 1 in \cite{mclain2013efficient} and is based on Theorem 3.1 of \cite{wang1985strong} on results for consistency of approximate maximum likelihood estimators. Note that $\hat{\bm\phi}_{N_0,N_1}=\hat{\bm\phi}_{\*N}=\left(\hat{\bm\beta}_{n, N_0,N_1},\hat{\beta}(\cdot)_{n, N_1}, \hat{H}_{N_0}\right)$ maximizes the log-likelihood $l_n(\bm\beta,\bm\gamma,\bm\theta|\*X,X_{f}(\cdot),\*Y)$
in equation (10) of the paper and is an approximate maximum likelihood estimator based on the Bernstein-Weierstrass approximation theorem \citep{lorentz2013bernstein}. Since, $N_n=O\left(n^{\kappa}\right)$, the rate at which $N_0,N_1$ increase must be such that $N_\ell \rightarrow \infty$ and $N_\ell / n \rightarrow$ 0 as $n \rightarrow \infty$ ($\ell=0,1$) and $\hat{\bm\phi}_{\*N}$ maximizes the log-likelihood as $N_0,N_1\rightarrow \infty$.

It is enough to show that \citep{mclain2013efficient} (suppressing the dependence of $\hat{\bm\beta}$ and $\hat{\beta}(s)$ on $N_0,N_1$)
\begin{eqnarray}
\int_{-\infty}^{\infty}\left|F_{\epsilon}\left\{\hat{H}_{N_0}(u)+\hat{\bm\beta}^T\*X+\int_{0}^{1} X_{f}(s)\hat{\beta}(s)ds\right\}-F_{\epsilon}\left\{H_{0}(u)+{\bm\beta}^T\*X+\int_{0}^{1} X_{f}(s){\beta}(s)ds\right\}\right | \notag\\ d F_{Y}(u \mid \*X,X_f(\cdot)) \rightarrow 0,
\end{eqnarray}
almost surely as $n \rightarrow \infty$. Theorem S1 then follows from the boundedness of the transformation function $H_{0}(\cdot)$ on $(0, \tau]$, the boundedness of $X_f(s)$, the continuity of the error c.d.f. $F_{\epsilon}$, and the dominated convergence theorem.

To show condition (1) we verify the five required assumptions of Theorem 3.1 in \cite{wang1985strong}.
Let us define $\pi\left(\hat{\phi}_{\*N}, t, \*X,X_f(\cdot)\right)=F_{\epsilon}\left\{\hat{H}_{N_0}(t)+\hat{\bm\beta}^T\*X+\int_{0}^{1} X_{f}(s)\hat{\beta}(s)ds\right\}$, with $\pi\left(\phi_{0}, t,\*X,X_f(\cdot) \right)=F_{T}(t \mid \*X,X_f(\cdot))$ (see equation (2) of the paper).

The assumption (i) in \cite{wang1985strong} is satisfied based on the definition of our metric, the assumption that $B$ is a bounded subset of $\mathbb{R}^p$ and since $\beta(\cdot)\in \mathcal{K}\subset\mathcal{L}^2[0,1]$. Denote $\Phi=B\times\mathcal{K}\times\mathcal{F}$. Then $(\Phi, d)$ is a separable compact metric space. Define $V_{b}\left(\phi_{0}\right), b \geq 1$ as a decreasing sequence of basic neighborhoods of $\pi\left(\phi_{0}, t,\*X,X_f(\cdot)\right)$ with radius $b^{-1}$. Let us also define $A_{\varphi}(\phi)$ as:
$
A_{\varphi}(\phi)=\varphi \pi(\phi, t, \*X,X_f(\cdot))+(1-\varphi) \pi\left(\phi_{0}, t, \*X,X_f(\cdot)\right).
$

Denote the observed data for subject i as $ \*Z_i=(Y_i,\Delta_i, \*X_i, X_{fi}(\cdot))$.
For all $\phi \in \Phi$, let $l\left\{\*{Z}, A_{\varphi}(\phi)\right\}=\Delta_{i} \log \left\{A_{\varphi}^{\prime}(\phi)\right\}+\left(1-\Delta_{i}\right) \log \left\{A_{\varphi}(\phi)\right\}$ with $A_{\varphi}^{\prime}(\phi)=$ $\partial A_{\varphi}(\phi) / \partial t$. Note that for every $b \geq 1$ there exists a $\varphi \in(0,1]$ with $\varphi \leq b^{-1}$ such that $A_{\varphi}(\phi) \in V_{b}\left(\phi_{0}\right)$. 
Under assumptions V and VI, to verify conditions (ii) and (iii) in \cite{wang1985strong} it is enough to show that (define $0 / 0 \equiv 0$)

\begin{equation}
E_{(Y, \Delta)}\left[l\left\{\*{Z}, A_{\varphi}(\phi)\right\} / l(\*{Z}, \phi) \mid \*X,X_f(\cdot)\right]>0, 
\end{equation}

The marginal density of $Y$ is given by

$$
f_{Y}(y \mid \*X,X_f(\cdot))=S_{C}(y) f_{T}(y \mid \*X,X_f(\cdot))+S_{T}(y) f_{C}(y \mid \*X,X_f(\cdot)) .
$$

Denoting $\pi^{\prime}=(\partial / \partial t) \pi$, and using $\pi\left(\phi_{0}, t,\*X,X_f(\cdot) \right)=F_{T}(t \mid \*X,X_f(\cdot))$, we have,

\begin{eqnarray}
   \int \frac{\pi^{\prime}(\phi, u, \*X,X_f(\cdot))}{\pi^{\prime}\left(\phi_{0}, u, \*X,X_f(\cdot)\right)} S_{C}(u) f_{T}(u \mid \*X,X_f(\cdot)) d u=\int \pi^{\prime}(\phi, u, \*X,X_f(\cdot)) S_{C}(u) d u \notag \\<\int \pi^{\prime}(\phi, u, \*X,X_f(\cdot)) d u \leq 1. \notag 
\end{eqnarray}
and
\begin{eqnarray}
    \int \frac{\pi(\phi, u,\*X,X_f(\cdot) )}{\pi\left(\phi_{0}, u, \*X,X_f(\cdot)\right)} S_{T}(u) f_{C}(u \mid \*X,X_f(\cdot)) d u=\int \pi(\phi, u, \*X,X_f(\cdot)) f_{C}(u) d u<\int f_{C}(u) d u=1 .\notag
\end{eqnarray}

Applying Jensen's inequality, we have

\begin{equation}
    \int \log \left\{\frac{\pi^{\prime}(\phi, u, \*X,X_f(\cdot))}{\pi^{\prime}\left(\phi_{0}, u, \*X,X_f(\cdot)\right)}\right\} S_{C}(u) f_{T}(u \mid \*X,X_f(\cdot)) d u<0.
\end{equation}
and

\begin{equation}
    \int \log \left\{\frac{\pi(\phi, u, \*X,X_f(\cdot))}{\pi\left(\phi_{0}, u, \*X,X_f(\cdot)\right)}\right\} S_{T}(u) f_{C}(u \mid \*X,X_f(\cdot)) d u<0.
\end{equation}
Hence based on (3) we have,

$$
\begin{aligned}
E_{Y}\left[\left.\frac{l\left(Y, 1, \*X,X_f(\cdot) , A_{\varphi}(\phi)\right)}{l\{Y, 1, \*X,X_f(\cdot) , \phi\}} \right\rvert\, \*X,X_f(\cdot) , \Delta=1\right] & =E_{Y}\left[\left.\log \left\{\frac{A_{\varphi}(\phi)}{\pi^{\prime}(\phi, Y, \*X,X_f(\cdot))}\right\} \right\rvert\, \*X,X_f(\cdot), \Delta=1\right] \\
& \geq E_{Y}\left[\left.\log \left\{\frac{\pi^{\prime}\left(\theta_{0}, Y, \*X,X_f(\cdot)\right)}{\pi^{\prime}(\theta, Y, \*X,X_f(\cdot))}\right\} \right\rvert\, \*X,X_f(\cdot), \Delta=1\right] \\
& >0,
\end{aligned}
$$
and similar result for $\Delta=0$ using (4). This verifies the condition (2). The last two conditions (iv) and (v) of \cite{wang1985strong} are easily verified by the continuity of $l\{\*{Z}, \phi\}$ and this proves the Theorem S1.

%\section{Appendix B: Asymptotic Normality of the Estimators}

%\section{Appendix C: Additional Simulations for Model selection performance using AIC}

\section{Appendix B: Supplementary Tables}
\begin{table}[H]
\centering
\caption{Estimated coverage from the $95\%$ Wald confidence intervals for the functional and scalar coefficients, Scenario A1. For the functional coefficients the average of the point-wise coverage over $S$ (grid of length $m=101$ on $\mathcal{S}=[0,1]$) is reported.}
\vspace{4 mm}
\label{tab:my-table2cov}
\begin{tabular}{llll}
\hline
Sample Size  & $\beta(s)$ & $\beta_1$ & $\beta_2$ \\ \hline
$n=100$     & 0.95   & 0.93       & 0.93  \\ \hline
$n=300$     & 0.97   & 0.92       & 0.94 \\ \hline
$n=500$    & 0.93   & 0.92       & 0.95  \\ \hline
\end{tabular}
\end{table}

\begin{table}[H]
\centering
\caption{Estimated coverage from the $95\%$ Wald confidence intervals for the functional and scalar coefficients, Scenario A2. For the functional coefficients the average of the point-wise coverage over $S$ (grid of length $m=101$ on $\mathcal{S}=[0,1]$) is reported.}
\vspace{4 mm}
\label{tab:my-table2cov}
\begin{tabular}{llll}
\hline
Sample Size  & $\beta(s)$ & $\beta_1$ & $\beta_2$ \\ \hline
$n=100$     & 0.94   & 0.93       & 0.95  \\ \hline
$n=300$     & 0.92   & 0.95       & 0.93 \\ \hline
$n=500$    & 0.97   & 0.96       & 0.98  \\ \hline
\end{tabular}
\end{table}

\begin{table}[H]
\centering
\caption{Descriptive Statistics of the Sample considered in the NHNAES 2011-2014 application. The mean (standard deviation) and count (percentage) is presented for continuous and categorical variables respectively.}
\label{tab:risk_factors}
\begin{tabular}{@{}lllll@{}}
\toprule
\textbf{Variable} & \textbf{Total (n=3032)} & \textbf{Deceased (n=582)} & \textbf{Survivors (n=2450)} & \textbf{p-value} \\ \midrule
Age  & 66.0 (9.36) & 72.31 (8.80) & 64.50 (8.85) & $<0.001$ \\
Gender & & & & $<0.001$ \\
\quad Female & 1687 (55.6) & 278 (47.8) & 1409 (57.5) \\
\quad Male & 1345 (44.4) & 304 (52.2) & 1041 (42.5) \\
BMI & 30.06 (7.07) & 28.89 (7.35) & 30.34 (6.98) & $<0.001$ 
\\ \bottomrule
\end{tabular}
\end{table}

\begin{table}[H]
\centering
\caption{Estimated Scalar coefficients from the FTTM method along with their $95\%$ confidence intervals in the NHANES application.}
\vspace{2 mm}
\label{tab:coefuci}
\begin{tabular}{l|rrrrrrr}
  \hline
 Variable &  Estimate (CI)  \\ 
  \hline
Age & 0.077 (0.015,0.139) \\ 
  Gender (Female)  & -0.129 (-0.457,0.198)   \\ 
  BMI  & -0.024 (-0.070,0.022) \\ 
   \hline
\end{tabular}
\vspace{0.5cm}
\end{table}

\begin{table}[H]
\centering
\caption{Descriptive Statistics of the Sample considered in the CGM application. Adverse Event  is defined as having 65 episodes of hypoglycemia during followup. The mean (standard deviation) and count (percentage) is presented for continuous and categorical variables respectively.}
\label{tab:risk_factors}
\begin{tabular}{@{}lllll@{}}
\toprule
\textbf{Variable} & \textbf{Total (n=296)} & \textbf{Adverse Event  (n=266)} & \textbf{No Adverse event (n=30)} & \textbf{p-value} \\ \midrule
Age  & 22.92 (14.68) & 23.5 (14.83) & 17.9 (12.35) & $ 0.048$ \\
Gender & & & & $0.43$ \\
\quad Female & 163 (55.1) & 149 (56.0) & 14 (46.67) \\
\quad Male & 133 (44.9) & 117 (44.0) & 16 (53.33) \\
HbA1C & 7.91 (0.64) & 7.87 (0.61) & 8.29 (0.77) & $<0.001$ 
\\ \bottomrule
\end{tabular}
\end{table}

\begin{table}[H]
\centering
\caption{Estimated Scalar coefficients from the FTTM method along with their $95\%$ confidence intervals in the CGM application.}
\vspace{2 mm}
\label{tab:coefuci}
\begin{tabular}{l|rrrrrrr}
  \hline
 Variable &  Estimate (CI)  \\ 
  \hline
Age & 0.009 (-0.027,0.046) \\ 
  Gender (Male)  & -0.492 (-1.22,0.239)   \\ 
  \hline
\end{tabular}
\vspace{0.5cm}
\end{table}

\section{Appendix C: Supplementary Figures}

\begin{figure}[H]
\centering
\includegraphics[width=0.9\linewidth , height=0.9\linewidth]{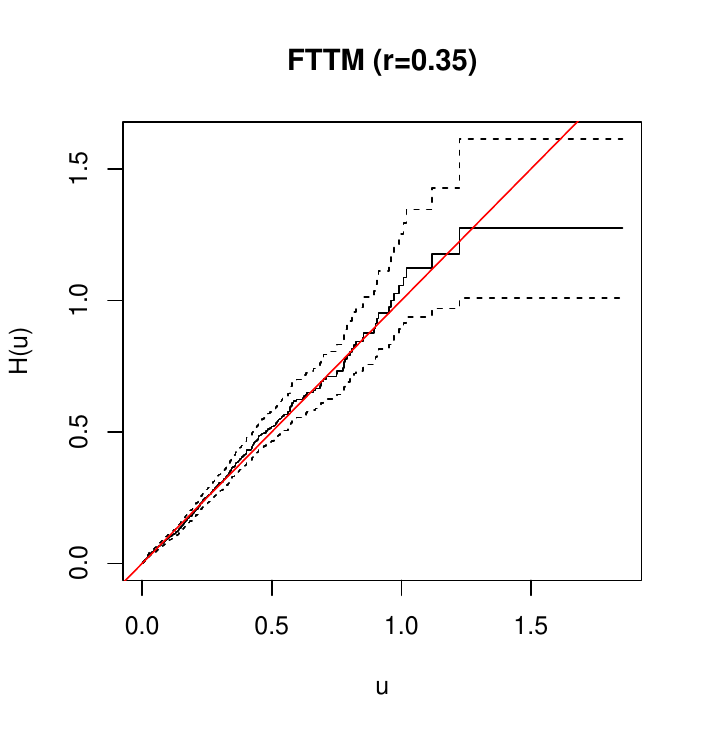} 
\caption{Nelson Allen estimate of cumulative hazard $\hat{\Lambda}(u)$ (solid line) and the associated $95\%$ confidence interval based on the fitted FTTM pseudo residuals $U_i$ for the NHANES 2011-2014 application.}
\label{fig:fig5s1}
\end{figure}

\begin{figure}[H]
\begin{center}
\begin{tabular}{ll}
\includegraphics[width=.5\linewidth , height=.5\linewidth]{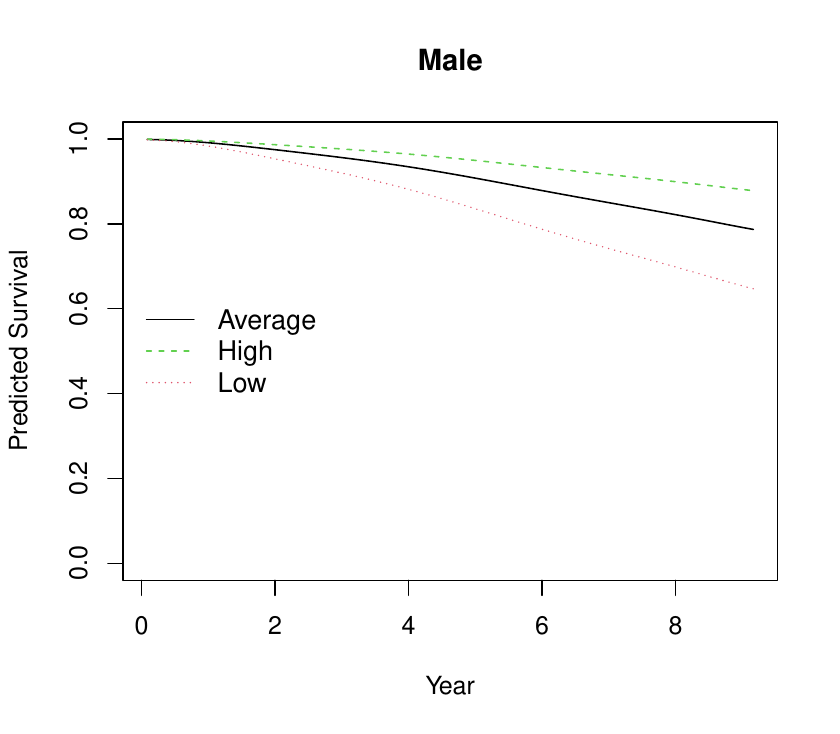} &
\includegraphics[width=.5\linewidth , height=.5\linewidth]{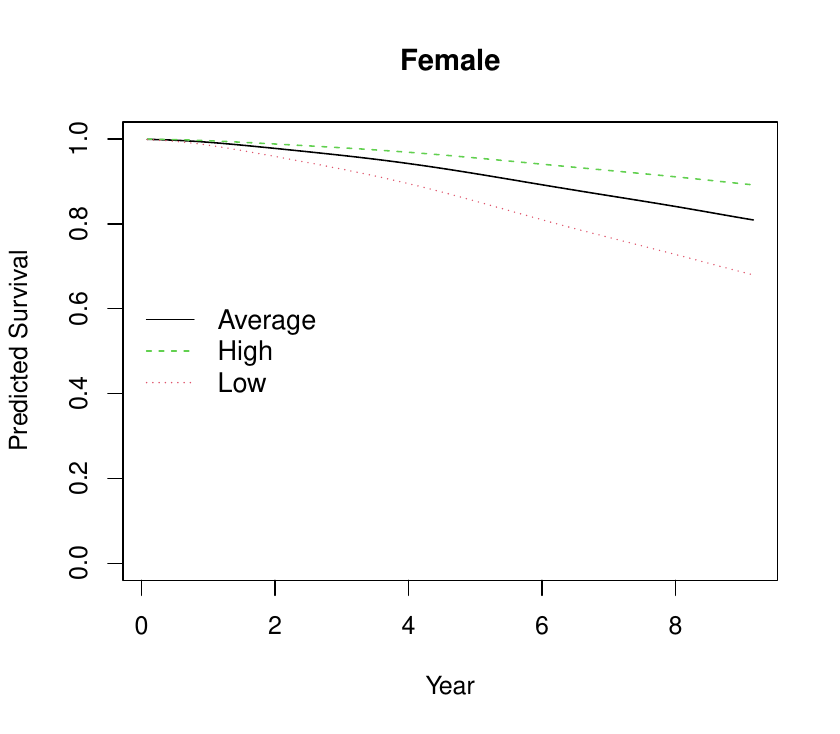}
\end{tabular}
\end{center}
\caption{Predicted survival probabilities from the FTTM for Male and Female participants with Age=66, BMI=30.06 and diurnal PA belonging to one of Average, High or Low group in the NHANES application. }
\label{fig:fig6s1}
\end{figure}

\begin{figure}[H]
\centering
\includegraphics[width=0.9\linewidth , height=0.9\linewidth]{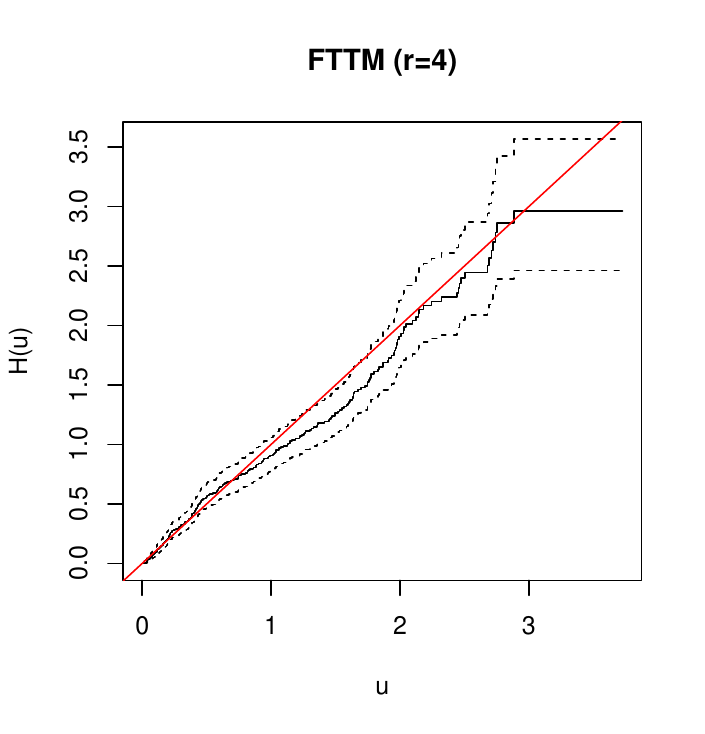} 
\caption{Nelson Allen estimate of cumulative hazard $\hat{\Lambda}(u)$ (solid line) and the associated $95\%$ confidence interval based on the fitted FTTM pseudo residuals $U_i$ for the CGM application.}
\label{fig:fig3s1}
\end{figure}

\bibliographystyle{Chicago}
\bibliography{refs}